\RequirePackage{fix-cm}
\documentclass[copyright,creativecommons]{eptcs}
\providecommand{\event}{Dave Schmidt's Festschrift}
\newcommand{\authorrunning}{B.-Y.~E.~Chang and X.~Rival}
\newcommand{\titlerunning}{Modular Construction of Shape-Numeric Analyzers}
\usepackage{cite} 
\input{packages.tex}

\newcommand{\mathboldcommand}[1]{\mathbb{#1}}
\newcommand{\bbA}{\mathboldcommand{A}}
\newcommand{\bbB}{\mathboldcommand{B}}
\newcommand{\bbC}{\mathboldcommand{C}}
\newcommand{\bbD}{\mathboldcommand{D}}
\newcommand{\bbE}{\mathboldcommand{E}}
\newcommand{\bbF}{\mathboldcommand{F}}

\newcommand{\bbH}{\mathboldcommand{H}}

\newcommand{\bbL}{\mathboldcommand{L}}
\newcommand{\bbM}{\mathboldcommand{M}}
\newcommand{\bbN}{\mathboldcommand{N}}
\newcommand{\bbO}{\mathboldcommand{O}}

\newcommand{\bbS}{\mathboldcommand{S}}
\newcommand{\bbT}{\mathboldcommand{T}}

\newcommand{\bbV}{\mathboldcommand{V}}

\newcommand{\bbX}{\mathboldcommand{X}}

\usepackage{euscript}
\newcommand{\mathcalcommand}[1]{\mathcal{#1}}

\newcommand{\mcC}{\mathcalcommand{C}}

\newcommand{\mcE}{\mathcalcommand{E}}

\newcommand{\mcL}{\mathcalcommand{L}}

\newcommand{\mcP}{\mathcalcommand{P}}

\DeclareMathAlphabet{\mathpzc}{T1}{pzc}{m}{it}

\newcommand{\mathfrakcommand}[1]{\mathfrak{#1}}

\newcommand{\fkU}{\mathfrakcommand{U}}

\newrgbcolor{mblue}{        0.1   0.1   0.5 }
\newrgbcolor{mbrown}{       0.57  0.4   0.3 }
\newrgbcolor{mcyan}{        0.1   0.1   0.5 }
\newrgbcolor{mgreen}{       0.2   0.5   0.2 }
\newrgbcolor{mmidgray}{     0.65  0.65  0.65}
\newrgbcolor{morange}{      0.8   0.45  0.45}
\newrgbcolor{mpurple}{      0.5   0.1   0.5 }
\newrgbcolor{mred}{         1     0     0   }
\newrgbcolor{mhigray}{      0.21  0.21  0.21}
\newrgbcolor{mgray}{        0.5   0.5   0.5 }
\newrgbcolor{mmidgreen}{    0.3   0.8   0.4 }
\newrgbcolor{mlightgray}{   0.85  0.85  0.85}
\newrgbcolor{mlightblue}{   0.80  0.90  1.00}
\newrgbcolor{mlightgreen}{  0.80  1.00  0.90}
\newrgbcolor{mlightpurple}{ 1.00  0.7   1.00}
\newrgbcolor{mlightred} {   1.00  0.7   0.7 }
\newrgbcolor{mlightyellow}{ 1.00  1.00  0.6 }
\newrgbcolor{myellow}{      1.00  1.00  0.30}

\newcommand{\eg}{e.g.\xspace}
\newcommand{\ie}{i.e.\xspace}

\newcommand{\apron}{\textsc{Apron}\xspace}

\newcommand{\partsof}[1]{\mcP(#1)}
\newcommand{\partsoffin}[1]{\mcP_{\rm fin}(#1)}
\newcommand{\subst}[3]{#1[#2 \leftarrow #3]}
\newcommand{\abs}[1]{{#1}^{\sharp}}
\newcommand{\trfin}[1]{{#1}^{\star}}
\newcommand{\sem}[1]{\llbracket #1 \rrbracket}
\newcommand{\asem}[1]{\sem{#1}^{\sharp}}
\newcommand{\lfp}{\textbf{lfp}}
\newcommand{\oplfp}[1][]{\mathop{\textbf{lfp}_{#1}}}

\newcommand{\alfp}{\abs{\lfp}}
\newcommand{\opalfp}[1][]{\mathop{\abs{\lfp}_{#1}}}
\newcommand{\sbools}{\bbB}
\newcommand{\false}{\textbf{false}}
\newcommand{\true}{\textbf{true}}
\newcommand{\undef}{\fkU}
\newcommand{\undefset}[1]{#1_{\fkU}}
\newcommand{\defeq}{\ensuremath{\mathrel{\smash{\stackrel{\mbox{\normalfont\tiny def}}{=}}}}}
\newcommand{\finitemap}{\rightharpoonup_{\text{\rm fin}}}

\newcommand{\svars}{\bbX}
\newcommand{\val}{v}
\newcommand{\svals}{\bbV}
\newcommand{\addr}{a}
\newcommand{\saddrs}{\bbA}
\newcommand{\heap}{\sigma}
\newcommand{\sheaps}{\bbH}
\newcommand{\env}{E}
\newcommand{\senvs}{\bbE}

\newcommand{\mem}{m}
\newcommand{\smems}{\bbM}
\newcommand{\heapdom}{\textbf{dom}}
\newcommand{\heapsubst}[3]{\subst{#1}{#2}{#3}}
\newcommand{\ctrl}{\ell}
\newcommand{\ctrlbeg}{\ell_{\mathrm{pre}}}
\newcommand{\ctrlend}{\ell_{\mathrm{post}}}
\newcommand{\sctrls}{\bbL}
\newcommand{\state}{s}
\newcommand{\States}{S}
\newcommand{\sstates}{\bbS}

\newcommand{\straces}{\bbT}
\newcommand{\prog}{p}
\newcommand{\lval}{\mathit{loc}}
\newcommand{\lvalsem}[1]{\mathord{\lvals{}}\sem{#1}}
\newcommand{\expr}{\mathit{exp}}
\newcommand{\exprsem}[1]{\mathord{\exprs{}}\sem{#1}}
\newcommand{\var}{x}
\newcommand{\rtrans}{\rightarrow}
\newcommand{\rtransp}{\rightarrow_{\prog}}
\newcommand{\rtranspstar}{\rightarrow^{\star}_{\prog}}
\newcommand{\semr}[1]{\sem{#1}_{\bf r}}
\newcommand{\semt}[1]{\sem{#1}_{\bf t}}
\newcommand{\semd}[1]{\sem{#1}_{\bf d}}
\newcommand{\semdh}[1]{\sem{#1}_{\bf dh}}
\newcommand{\Fsem}{F}
\newcommand{\Fsemr}{\Fsem_{\bf r}}

\newcommand{\emp}{{\bf emp}}
\newcommand{\aster}{\mathbin{\mbox{\relsize{1}\raisebox{-0.1ex}{$\ast$}}}}
\newcommand{\lsep}{\mathrel{\mathord{\aster}}}

\newcommand{\avars}{\abs{\svals}}
\newcommand{\valua}{\nu}
\newcommand{\gvars}[1]{\abs{\svals}{[#1]}}
\newcommand{\relpt}[1]{\stackrel{#1}{\mapsto}}
\newcommand{\pt}{\relpt{}}

\newcommand{\icallpz}[2]{#1 \cdot #2}

\newcommand{\inddef}[1]{\mathbf{#1}}

\newcommand{\indlist}{\inddef{list}}
\newcommand{\relunfold}{\leadsto}

\newcommand{\aF}{\abs{F}}
\newcommand{\af}{\abs{f}}
\newcommand{\ax}{\abs{d}}
\newcommand{\cset}{\bbO}
\newcommand{\cdom}{\bbD}
\newcommand{\adom}{\bbD^{\sharp}}
\newcommand{\aheaps}{\abs{\sheaps}}
\newcommand{\gammaheap}{\gamma_{\sheaps}}
\newcommand{\aheap}{\abs{\heap}}
\newcommand{\adomnump}[1]{\bbD_{\rm num} \langle #1 \rangle}
\newcommand{\gammanump}[1]{\gamma_{\rm num} \langle #1 \rangle}
\newcommand{\adomnum}{\abs{\bbN}}
\newcommand{\anum}{\abs{\valua}}
\newcommand{\adomcomb}{\aheaps \rightrightarrows \adomnum}
\newcommand{\gammacomb}{\gamma_{\aheaps \rightrightarrows \adomnum}}
\newcommand{\symconvsym}{\Phi}
\newcommand{\symconv}[1]{\symconvsym \langle #1 \rangle}
\newcommand{\adommem}{\abs{\smems}}
\newcommand{\amems}{\adommem}
\newcommand{\amem}{\abs{\mem}}
\newcommand{\gammamem}{\gamma_{\smems}}
\newcommand{\aenv}{\abs{\env}}
\newcommand{\aenvs}{\abs{\senvs}}
\newcommand{\adomdisj}{\abs{\smems}_{\vee}}
\newcommand{\gammadisj}{\gamma_{\vee}}
\newcommand{\astate}{\abs{\state}} 
\newcommand{\aState}{\abs{S}}      

\newcommand{\upgrade}[2]{#2_{\rm #1}}
\newcommand{\abscommand}[1]{\mathfrak{#1}}
\newcommand{\absevall}{\abscommand{eval[l]}}
\newcommand{\absevale}{\abscommand{eval[e]}}
\newcommand{\absassign}{\abscommand{assign}}
\newcommand{\absguard}{\abscommand{guard}}
\newcommand{\absmutate}{\abscommand{mutate}}
\newcommand{\absnew}{\abscommand{new}}
\newcommand{\absdeleten}{\abscommand{delete[n]}}
\newcommand{\absdeletee}{\abscommand{delete[e]}}
\newcommand{\absalloc}{\abscommand{alloc}}
\newcommand{\absfree}{\abscommand{free}}
\newcommand{\absrename}{\abscommand{rename}}
\newcommand{\absunfold}{\abscommand{unfold}}
\newcommand{\abscompare}{\abscommand{compare}}
\newcommand{\absjoin}{\abscommand{join}}
\newcommand{\abswiden}{\abscommand{widen}}
\newcommand{\abspartition}{\abscommand{partition}}
\newcommand{\abscollapse}{\abscommand{collapse}}
\newcommand{\upgradeheap}{\upgrade{shape}}
\newcommand{\evallheap}{\upgradeheap{\absevall}}
\newcommand{\evaleheap}{\upgradeheap{\absevale}}
\newcommand{\unfoldheap}{\upgradeheap{\absunfold}}
\newcommand{\mutateheap}{\upgradeheap{\absmutate}}
\newcommand{\newheap}{\upgradeheap{\absnew}}
\newcommand{\deletenheap}{\upgradeheap{\absdeleten}}
\newcommand{\deleteeheap}{\upgradeheap{\absdeletee}}
\newcommand{\compareheap}{\upgradeheap{\abscompare}}
\newcommand{\joinheap}{\upgradeheap{\absjoin}}
\newcommand{\widenheap}{\upgradeheap{\abswiden}}
\newcommand{\upgradenum}{\upgrade{num}}
\newcommand{\assignnum}{\upgradenum{\absassign}}
\newcommand{\guardnum}{\upgradenum{\absguard}}
\newcommand{\newnum}{\upgradenum{\absnew}}
\newcommand{\deletennum}{\upgradenum{\absdeleten}}
\newcommand{\renamenum}{\upgradenum{\absrename}}
\newcommand{\comparenum}{\upgradenum{\abscompare}}
\newcommand{\joinnum}{\upgradenum{\absjoin}}
\newcommand{\widennum}{\upgradenum{\abswiden}}
\newcommand{\upgradecomb}{\upgrade{comb}}
\newcommand{\unfoldcomb}{\upgradecomb{\absunfold}}
\newcommand{\assigncomb}{\upgradecomb{\absassign}}
\newcommand{\guardcomb}{\upgradecomb{\absguard}}
\newcommand{\alloccomb}{\upgradecomb{\absalloc}}
\newcommand{\freecomb}{\upgradecomb{\absfree}}
\newcommand{\comparecomb}{\upgradecomb{\abscompare}}
\newcommand{\joincomb}{\upgradecomb{\absjoin}}
\newcommand{\widencomb}{\upgradecomb{\abswiden}}
\newcommand{\upgrademem}{\upgrade{mem}}
\newcommand{\assignmem}{\upgrademem{\absassign}}
\newcommand{\guardmem}{\upgrademem{\absguard}}
\newcommand{\allocmem}{\upgrademem{\absalloc}}
\newcommand{\freemem}{\upgrademem{\absfree}}
\newcommand{\comparemem}{\upgrademem{\abscompare}}
\newcommand{\joinmem}{\upgrademem{\absjoin}}
\newcommand{\widenmem}{\upgrademem{\abswiden}}
\newcommand{\upgradedisj}{\upgrade{\vee}}
\newcommand{\partitiondisj}{\upgradedisj{\abspartition}}
\newcommand{\collapsedisj}{\upgradedisj{\abscollapse}}
\newcommand{\assigndisj}{\upgradedisj{\absassign}}
\newcommand{\guarddisj}{\upgradedisj{\absguard}}
\newcommand{\allocdisj}{\upgradedisj{\absalloc}}
\newcommand{\freedisj}{\upgradedisj{\absfree}}
\newcommand{\comparedisj}{\upgradedisj{\abscompare}}
\newcommand{\joindisj}{\upgradedisj{\absjoin}}
\newcommand{\widendisj}{\upgradedisj{\abswiden}}
\newcommand{\ctxts}{\bbC}
\newcommand{\ctxt}{c}
\newcommand{\ctxtof}[1]{\mcC[#1]}
\newcommand{\relcw}{\Phi}
\newcommand{\reljoin}{\Psi}

\newcommand{\lvals}[1]{\mcL_{#1}}
\newcommand{\exprs}[1]{\mcE_{#1}}
\newcommand{\progs}[1]{\mcP_{#1}}
\newcommand{\sfields}{\bbF}
\newcommand{\offzero}{\underline{\emptyset}}
\newcommand{\ccom}[1]{\textbf{#1}}

\newcommand{\celse}{\ccom{else}}

\newcommand{\cfree}{\ccom{free}}
\newcommand{\cif}{\ccom{if}}
\newcommand{\cmalloc}{\ccom{malloc}}

\newcommand{\cstruct}{\ccom{struct}}

\newcommand{\cunion}{\ccom{union}}

\newcommand{\cwhile}{\ccom{while}}

\newcommand{\caddressof}{\texttt{\&}}
\newcommand{\cderef}{\mathtt{\star}}
\newcommand{\cfield}{\mathbin{\texttt{->}}}
\newcommand{\cstar}{\star}
\newcommand{\ctyp}[1]{\textbf{#1}}

\newcommand{\cint}{\ctyp{int}}

\newcommand{\clist}{\ctyp{list}}

\newcommand{\ttvar}[1]{\mathtt{#1}}

\newcommand{\vars}{\ttvar{s}}

\newcommand{\varx}{\ttvar{x}}
\newcommand{\vary}{\ttvar{y}}

\newcommand{\makefield}[1]{\underline{\mathrm{#1}}}
\newcommand{\flda}{\makefield{a}}
\newcommand{\fldb}{\makefield{b}}
\newcommand{\fldc}{\makefield{c}}
\newcommand{\fldd}{\makefield{d}}
\newcommand{\fldf}{\makefield{f}}
\newcommand{\fldg}{\makefield{g}}
\newcommand{\fldn}{\makefield{next}}

\newcommand{\relatedworkpara}{\smallskip\emph{Related work and
    discussion}. }

\title{Modular Construction of Shape-Numeric Analyzers
}
\author{Bor-Yuh Evan Chang
\institute{University of Colorado Boulder}
\email{bec@cs.colorado.edu}
\and
Xavier Rival
\institute{INRIA, ENS, and CNRS}
\email{rival@di.ens.fr}
}
\begin{document}
\maketitle
\begin{abstract}
The aim of static analysis is to infer invariants about programs that are precise enough to establish semantic properties, such as the absence of run-time errors. Broadly speaking, there are two major branches of static analysis for imperative programs.  Pointer and \emph{shape} analyses focus on inferring properties of pointers, dynamically-allocated memory, and recursive data structures, while \emph{numeric} analyses seek to derive invariants on numeric values. Although simultaneous inference of shape-numeric invariants is often needed, this case is especially challenging and is not particularly well explored.  Notably, simultaneous shape-numeric inference raises complex issues in the design of the static analyzer itself.
%

In this paper, we study the construction of such shape-numeric, static analyzers. We set up an abstract interpretation framework that allows us to reason about simultaneous shape-numeric properties by combining shape and numeric abstractions into a modular, expressive abstract domain.  Such a modular structure is highly desirable to make its formalization and implementation easier to do and get correct. To achieve this, we choose a concrete semantics that can be abstracted step-by-step, while preserving a high level of expressiveness.  The structure of abstract operations (i.e., transfer, join, and comparison) follows the structure of this semantics.  The advantage of this construction is to divide the analyzer in modules and functors that implement abstractions of distinct features.
%
\end{abstract}

\section{Introduction}
\label{sec:1:intro}

The static analysis of programs written in real-world imperative
languages like C or Java are challenging because of the mix of
programming features that the analyzer must handle effectively. On one
hand, there are pointer values (i.e., memory addresses) that can be
used to create dynamically-allocated recursive data structures. On the
other hand, there are numeric data values (e.g., integer and
floating-point values) that are integral to the behavior of the
program. While it is desirable to use distinct abstract domains to
handle such different families of properties, precise analyses require
these abstract domains to {\em exchange} information because the
pointer and numeric values are often interdependent. Setting up the
structure of the implementation of such a shape-numeric analyzer can
be quite difficult. While maintaining separate modules with clearly
defined interfaces is a cornerstone of software engineering, such
boundaries also impede the easy exchange of semantic information.
%


In this manuscript, we contribute a modular construction of an
abstract domain~\cite{cc:popl:77} that layers a numeric abstraction on
a shape abstraction of memory.  The construction that we present is
parametric in the numeric abstraction, as well as the shape
abstraction.  For example, the numeric abstraction may be instantiated
with an abstract domain such such as polyhedra~\cite{ch:popl:78} or
octagons~\cite{am:hosc:06}, while the shape abstraction may be
instantiated with domains such as
Xisa~\cite{xisa:popl:08,bec:phd:2008} or TVLA~\cite{tvla:toplas:02}.
Note that the focus of this paper is on describing the formalization
and construction of the abstract domain.  Empirical evaluation of
implementations based on this construction are given
elsewhere~\cite{xisa:sas:07,bec:phd:2008,xisa:popl:08,xisa:esop:10,xisa:popl:11,sr:aplas:12,tcr:vmcai:13}.

We describe our construction in four steps:
\begin{compactenum}
\item We define a concrete program semantics for a generic imperative
  programming language focusing on the concrete model of mutable
  memory (Section~\ref{sec:2:conc}).
\item We describe a step-by-step abstraction of program states as a
  cofibered construction of a numeric abstraction layer on top of a
  shape abstraction layer (Section~\ref{sec:3:abs}).  In particular,
  we characterize a shape abstraction as a combination of an
  \emph{exact} abstraction of memory cells along with a
  \emph{summarization} operation.  Then, we describe how a value
  abstraction can be applied both globally on \emph{materialized}
  memory locations and locally within summarized regions.
\item We detail the abstract operators necessary to implement an
  abstract program semantics in terms of interfaces that a shape
  abstraction and a value abstraction must implement
  (Section~\ref{sec:4:ops}).
\item We overview a modular construction of a shape-numeric static
  analyzer based on our abstract operators (Section~\ref{sec:5:ai}).
\end{compactenum}

\section{A concrete semantics}
\label{sec:2:conc}


We first define a concrete program semantics for a generic imperative
programming language.

\subsection{Concrete memory states}

We define a ``bare metal'' model of machine memory.  A {\em concrete store} is
a partial function \( \heap \in \sheaps = \saddrs \finitemap \svals \) from addresses to values.
An address \( \addr \in \saddrs \) is also considered a value \( \val
\in \svals \), that is, we assume that \( \saddrs \subseteq \svals \).
For simplicity, we assume that all cells of any store \( \heap \) have
the same size (i.e., word-sized) and that all addresses are aligned
(i.e., word-aligned).  For example, we can imagine a standard 32-bit
architecture where all values are 4-bytes and all addresses are
4-byte--aligned.  We write for \( \heapdom( \heap ) \) the set of
addresses at which \( \heap \) is defined, and we let \(
\heapsubst{\heap}{\addr}{\val} \) denote the heap obtained after
updating the cell at address \( \addr \) with value \( \val \).
A {\em concrete environment} \( \env \in \senvs = \svars \rightarrow
\saddrs \) maps program variables to their addresses.  That is, we
consider all program variables as mutable cells in the concrete
store---the concrete environment $\env$ indicates where each variable
is allocated.  A {\em concrete memory state} \( \mem \) simply pairs a
concrete environment and a concrete store: \( (\env, \heap) \). Thus,
the set of memory states \( \smems = \senvs \times \sheaps \) is the
product of the set of concrete environments and the set of concrete
stores.

\begin{figure}[tb]\centering
    \includegraphics[scale=1]{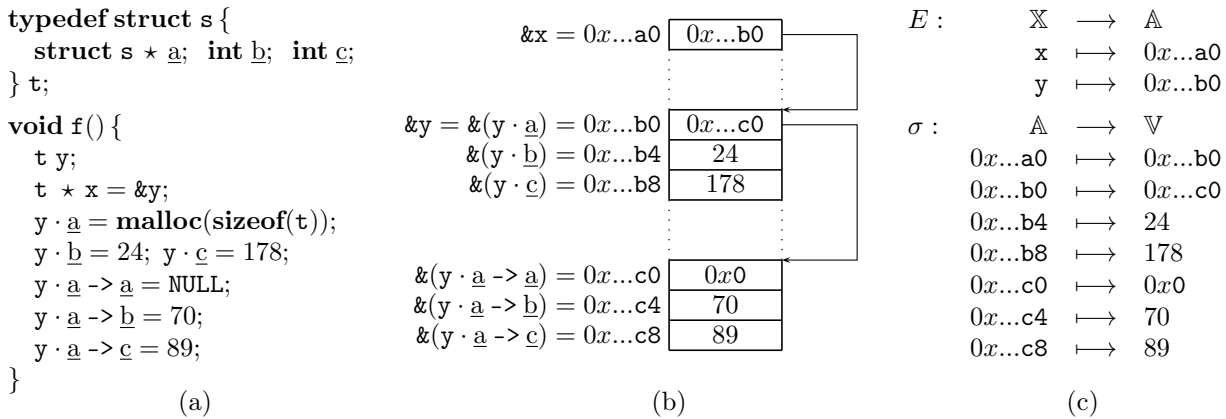}
  \caption{A concrete memory state consists of an environment $\env$
    and a store $\heap$ shown in (c).  This example state corresponds
    to the informal box diagram shown in (b) and a state at the 
    return point of the C-procedure $\mathtt{f}$ in (a).}
  \label{fig:1:exmem}
\end{figure}
Figure~\ref{fig:1:exmem}(c) shows an example concrete memory state at
the return point of the procedure $\mathtt{f}$ in (a).  The
environment $\env$ has two bindings for the variables $\varx$ and
$\vary$ that are in scope. For concreteness, we show the concrete
store for this example laid out using 32-bit addresses and a C-style
layout for $\cstruct \; \vars$.  The figure shown in (b) shows the
concrete store as an informal box diagram.

\relatedworkpara
Observe that we do not make the distinction between stack and heap
space in a concrete store $\heap$ (as in a C-style model), nor have we
partitioned a heap on field names (as in Java-style model).  We have
intentionally chosen this rather low-level definition of concrete
memory states---essentially an assembly-level model of memory---and
leave any abstraction to the definition of abstract memory states.  An
advantage of this approach is the ability to use a common concrete
model for combining abstractions that make different choices about the
details they wish to expose or hide~\cite{xisa:esop:10}.  For example,
Laviron et al.~\cite{xisa:esop:10} defines an abstract domain that
treats precisely C-style aggregates: both $\cstruct$s and $\cunion$s
with sized-fields and pointer arithmetic.  Another abstract
domain~\cite{sr:aplas:12} abstracts nested structures using a
hierarchical abstraction.  Rival and Chang~\cite{xisa:popl:11} defines
an abstraction that simultaneously summarizes the stack of activation
records and the heap data structures (with a slightly extended notion
of concrete environments), which is useful for analyzing recursive
procedures.

\subsection{Concrete program semantics}
\label{sec:2:1:concrete-semantics}

For the most part, we can be agnostic about the particulars of the
imperative programming language of interest.  To separate concerns
between abstracting memory and control points on which abstract
interpretation collects, all we assume is that a {\em concrete
  execution state} consists of a {\em control state} and a concrete
memory state.
A shape-numeric abstract domain as we define in Section~\ref{sec:3:abs}
abstracts the concrete memory state component.
\begin{definition}[Execution states]
  An {\em execution state} \( \state \in \sstates \) consists of a
  triple \( (\ctrl, \env, \heap) \) where \( \ctrl \in \sctrls \) is a
  control state, \( \env \in \senvs \) is an concrete environment, and
  \( \heap \in \sheaps \) is a concrete store.  The memory component
  of an execution state is the pair \( (\env, \heap) \in \smems \).
\end{definition}
\noindent
Thus, the set of execution states \( \sstates = \sctrls \times \senvs
\times \sheaps \equiv \sctrls \times \smems \).  A {\em program
  execution} is described by a {\em finite trace}, that is, a finite
sequence of states \( \langle \state_0, \ldots, \state_n \rangle \).
We let \( \straces = \trfin{\sstates} \) denote the set of finite
traces over \( \sstates \).


\begin{figure}
\[
\begin{array}{@{}ccc@{}}
  \begin{array}{rcll}
    \lval \; (\in \lvals{\svars})
    & ::=
    & \var
    & (\var \in \svars)
    \\
    & \mid
    & \lval_1 \cdot \fldf
    & (\lval_1 \in \lvals{\svars}; \fldf \in \sfields)
    \\
    & \mid
    & \cderef \expr
    & (\expr \in \exprs{\svars})
    \\
    \\
    \\
  \end{array}
  & &
  \begin{array}{rcll}
    \expr \; (\in \exprs{\svars})
    & ::=
    & \lval
    & (\lval \in \lvals{\svars})
    \\
    & \mid
    & \caddressof \lval
    & (\lval \in \lvals{\svars})
    \\
    & \mid
    & \val
    & (\val\in \svals)
    \\
    & \mid
    & \oplus(\overline{\expr})
    & (\overline{\expr} \in \exprs{\svars})
    \\
    \oplus 
    & ::=
    & \cdots
    \\
  \end{array} \\
\end{array}
\]
\[
\begin{array}{rclll}
  \prog \; (\in \progs{\svars})
  & ::=
  & \lval = \expr
  & \quad (\lval \in \lvals{\svars}; \expr \in \exprs{\svars}) \qquad
  & \text{assignment}
  \\
  & \mid
  & \lval = \cmalloc( \{ \fldf_1, \ldots, \fldf_n \} )
  & \quad (\lval \in \lvals{\svars}; [\fldf_1, \ldots, \fldf_n] \in \sfields^{\ast}) \qquad
  & \text{memory allocation}
  \\
  & \mid
  & \cfree( \lval )
  & \quad (\lval \in \lvals{\svars}) \qquad
  & \text{memory deallocation}
  \\
  & \mid
  & \prog_1; \prog_2
  & \quad (\prog_1, \prog_2 \in \progs{\svars}) \qquad
  & \text{sequence}
  \\
  & \mid
  & \cif \; ( \expr ) \; \prog_1 \; \celse \; \prog_1
  & \quad (\expr \in \exprs{\svars}; \prog_1, \prog_2 \in \progs{\svars}) \qquad
  & \text{condition test}
  \\
  & \mid
  & \cwhile \; ( \expr ) \; \prog_1
  & \quad (\expr \in \exprs{\svars}; \prog_1 \in \progs{\svars}) \qquad
  & \text{loop}
  \\
\end{array}
\]
\caption{Abstract syntax for a C-like imperative programming language.
  A program $\prog$ consists of assignment, dynamic memory allocation
  and deallocation, sequences, condition tests, and loops.  An
  assignment is specified by a location expression $\lval$ that names
  a memory cell to update and an expression $\expr$ that is evaluated
  to yield the new contents for the cell. For simplicity, we specify
  allocation with a list of field names (i.e., $\cmalloc(\{\fldf_1, \ldots, \fldf_n\})$).}
\label{fig:syntax}
\end{figure}

%

To make our examples more concrete, we consider a C-like programming
language whose syntax is shown in Figure~\ref{fig:syntax}.  A location
expression $\lval$ names a memory cell, which can be a program
variable $\var$, a field offset from another memory location $\lval_1
\cdot \fldf$, or the memory location named by a pointer value $\cderef
\expr$. We write \( \fldf \in \sfields \) for a field name and
implicitly read any field as an offset, that is, we write $\addr +
\fldf$ for the address $\addr' \in \saddrs$ obtained by offsetting an
address $\addr$ with field $\fldf$.  To emphasize that we mean C-style
field offset as opposed to Java-style field dereference, we write
$\varx \cdot \fldf$ for what is normally written as
$\varx\mathtt{.}\fldf$ in C.  As in C, we write $\expr \cfield \fldf$
for Java-style field dereference, which is a shorthand for $(\cderef
\expr) \cdot \fldf$.  An expression $\expr$ can be a memory location
expression $\lval$, an address of a memory location $\caddressof
\lval$, or any value literal $\val$, some other n-ary operator
$\oplus(\overline{\expr})$.  Like in C, a memory location expression
$\lval$ used as an expression (i.e., ``r-value'') refers to the contents
of the named memory cell, while the $\caddressof \lval$ converts the
location's address (i.e., ``l-value'') into a pointer ``r-value.''  We
leave the value literals $\val$ (e.g., $\mathtt{1}$) and expression
operators $\oplus$ (e.g., $!$, $+$, $==$) unspecified.  



\paragraph{An operational semantics:}

Given a program $\prog$, we assume its execution is described by a
transition relation \( \rtransp \subseteq \sstates \times \sstates \).
This relation defines a small-step operational semantics, which can be
defined as a structured operational semantics judgment $\state
\rtransp \state'$.  Such a definition is completely standard for our
language, so we do not detail it here.

\begin{wrapfigure}{r}{0pt}
\begin{math}
\begin{array}{@{}c@{}}
\inferrule[DoAssignment]{
}{
(\ctrlbeg, \env, \heap)
\rtrans_{\lval\,=\,\expr}
(\ctrlend, \env, \subst{\heap}{
  \lvalsem{\lval}( \env, \heap)}{\exprsem{\expr}( \env, \heap )})
}
\\[4ex]
\begin{array}{@{}r@{\;}c@{\;}l@{\qquad}r@{\;}c@{\;}l@{}}
  \lvalsem{\var}(\env, \heap) &\defeq& \env(\var)
  &
  \lvalsem{\cderef \expr} &\defeq& \exprsem{\expr}
  \\
  \exprsem{\lval}(\env, \heap) &\defeq& \heap \circ \lvalsem{\lval}(\env, \heap)
  &
  \exprsem{\caddressof \lval} &\defeq& \lvalsem{\lval}
  \\
\end{array}
\\[2.5ex]
\begin{array}{@{}r@{\;}c@{\;}l@{}}
  \lvalsem{\lval \cdot \fldf}(\env, \heap) &\defeq&
  \lvalsem{\lval}(\env, \heap) + \fldf
\end{array}
\end{array}
\end{math}
\caption{A small-step operational semantics for programs.}
\label{fig:opsem-assign}
\end{wrapfigure}
As an example rule, consider the case for an assignment \( \lval =
\expr \) where \( \ctrlbeg \) and \( \ctrlend \) are the control
points before and after the assignment, respectively.  We assume that
the semantics of a location expression \( \lvalsem{\lval} \) is a function
from memory states to addresses \( \smems \rightarrow \saddrs \) and
that the semantics of an expression \( \exprsem{\expr} \) is a function
from memory states to values \( \smems \rightarrow \svals \).  Then,
the transition relation for assignment simply updates the input store
$\heap$ at the address given by $\lval$ with the value given by
$\expr$ as shown in Figure~\ref{fig:opsem-assign}.
%
%
The evaluation of locations \( \lval \) and expressions \( \expr \),
that is, \( \lvalsem{\lval}( \env, \heap ) \) and \( \exprsem{\expr}(
\env, \heap ) \), respectively, can be defined by induction on their
structure.  The environment \( \env \) is used to lookup the allocated
address for program variables in $\lvalsem{\var}$. The value for a
memory location $\exprsem{\lval}$ is obtained by looking up the
contents in the store \( \heap \).  Dereference $\cderef \expr$ and
$\caddressof \lval$ mediate between address and value evaluation,
while field offset $\lval \cdot \fldf$ is simply an address
computation.  The evaluation of the remaining expression forms is
completely standard.

\begin{example}[Evaluating an assignment]
  Using the concrete memory state $(\env, \heap)$ shown in
  Figure~\ref{fig:1:exmem}, the evaluation of the assignment \( \varx
  \cfield \flda \cfield \fldb = \vary \cdot \fldc \) proceeds as
  follows.
First, the right-hand side
gets evaluated by noting that $\env(\vary) = 0x\mathtt{...b0}$ and
following
\[
  \exprsem{\vary \cdot \fldc}( \env, \heap )
  = \heap( \lvalsem{\vary \cdot \fldc}(\env, \heap) )
  = \heap( \lvalsem{\vary}(\env, \heap) + \fldc  )
  = \heap( \env(\vary) + \fldc  )
  = \heap( 0x\mathtt{...b8} )
  = 178 \;.
\]
Second, the left-hand side gets evaluated by noting that
$\env(\varx) = 0x\mathtt{...a0}$ and then following the location
evaluation
\(
  \lvalsem{\varx \cfield \flda \cfield \fldb}(\env, \heap)
  = \heap( \heap( 0x\mathtt{...a0} ) + \flda ) + \fldb )
  = \heap( 0x\mathtt{...b0} + \flda ) + \fldb
  = 0x\mathtt{...c0} + \fldb
  = 0x\mathtt{...c4}
\).  Finally, the store is updated at address $0x\mathtt{...c4}$ with
the value $178$ with \( \heapsubst{\heap}{ 0x\mathtt{...c4}}{178} \).
\end{example}


\paragraph{Concrete program semantical definitions:}

Several notions of program semantics can be used as a basis for static
analysis, which each depend on the desired properties and the kinds of
invariants needed to establish them.  A semantical definition
expressed as the least fixed-point of a continuous function $\Fsem$
over a concrete, complete lattice is particularly well-suited to the
design of abstract interpreters~\cite{cc:popl:77}.  Following this
analysis design methodology, an abstract interpretation consists of
(1) choosing an abstraction of the concrete lattice
(Section~\ref{sec:3:abs}), (2) designing abstract operators that
over-approximate the effect of the transition relation \( \rtransp \)
and concrete joins \( \cup \) (Section~\ref{sec:4:ops}), and (3)
applying abstract operators to over-approximate $\Fsem$ using widening
(Section~\ref{sec:5:ai}).

\begin{definition}[A concrete domain]
  Let us fix a form for our concrete domains $\cdom$ to be the
  powerset of some set of concrete objects $\cset$, that is, let
  $\cdom = \partsof{\cset}$.  Domain $\cdom$ form a complete lattice
  with subset containment $\subseteq$ as the partial order.  Hence,
  concrete joins are simply set union $\cup$.
\end{definition}

For a program \( \prog \), let \( \ctrlbeg \) be its entry point
(i.e., its initial control state).  A standard definition of interest
is the set of reachable states, which is sufficient for reasoning
about safety properties.
\begin{example}[Reachable states]
We write $\semr{\prog}$ for the set of reachable states of program
$\prog$, that is,
\[
  \semr{\prog} \defeq
  \{ \state \mid
  \text{$(\ctrlbeg, \env, \heap) \rtranspstar \state$
  for some $\env \in \senvs$ and $\heap \in \sheaps$}
  \} 
\]
where $\rtranspstar$ is the reflexive-transitive closure of the
single-step transition relation $\rtrans$.  Alternatively,
$\semr{\prog}$ can be defined as $\oplfp \Fsemr$, the least-fixed
point of $\Fsemr$, where $\Fsemr
: \partsof{\sstates} \rightarrow \partsof{\sstates}$ is as follows:
\[
\Fsemr(\States) \defeq
\{
  (\ctrlbeg, \env, \heap)
  \mid \text{$\env \in \senvs$ and $\heap \in \sheaps$}
\}
\cup
\{ \state' \mid \text{$\state \in \States$ and $\state \rtransp
  \state'$ for some $\state' \in \sstates$} \}
\;.
\]
Note that we have let the concrete objects $\cset$ be the execution
states $\sstates$ in this example.

We can also describe the reachable states
\emph{denotationally}~\cite{ds:den:09}---\(
 \semd{\prog}( \env, \heap ) \defeq \{ \state \mid
  \text{$(\ctrlbeg, \env, \heap) \rtranspstar \state$}
  \} 
\)---that
enables a compositional way to reason about programs.  Here, we
let the set of concrete objects be functions from memory states to
sets of states (i.e., \( \smems \rightarrow \partsof{\sstates} \)).
\end{example}

\relatedworkpara
For additional precision or for richer properties, it may be critical
to retain some information about the history of program executions
(i.e., how a state can be reached)~\cite{rm:toplas:07}.  In this case,
we might choose a {\em trace semantics} as a concrete semantics where
the concrete objects $\cset$ are chosen to be traces $\straces$.  For
instance, the finite prefix traces semantics is defined by \(
\semt{\prog} \defeq \{ \langle \state_0, \ldots, \state_n \rangle \mid
\state_0\colon (\ctrlbeg, \env_0, \heap_0) \;\text{and}\; \state_{i}
\rtransp \state_{i+1} \;\text{for some $\env_0 \in \senvs, \heap_0 \in
  \sheaps$ and for all $0 \leq i < n$} \} \).
%
Or we may to choose to define a trace semantics denotationally
\( \semdh{\prog}: \smems \rightarrow \partsof{\straces} \) that maps
input memory states into traces starting from them.

In this section, we have left the definition of a control state
essentially abstract.  A control state is simply a member of a set of
labels on which an interpreter visits.  In the intraprocedural
setting, the control state is usually a point in the program text
corresponding to a program counter.  Since the set of program points
is finite, the control state can be left unabstracted yielding a
flow-sensitive analysis.  Meanwhile, richer notions of control states
are often needed for interprocedural
analysis~\cite{mrr:tosem:05,sp:81}.

\section{Abstraction of memory states}
\label{sec:3:abs}
In this section, we discuss the abstraction of memory states,
including environments and stores, as well as the values stored in
them.  A shape abstraction typically abstracts entire stores but only
the pointer values (i.e., addresses) in them.  In contrast, a numeric
abstraction is typically applied only to the data values stored in
program variables (i.e., the part of the store containing the global
and local variables).  We defer the abstraction of program executions
to Section~\ref{sec:5:ai}.

Following the abstract interpretation framework~\cite{cc:popl:77}, an
{\em abstraction} or {\em abstract domain} is a set of abstract
properties \( \adom \) together with a concretization function and
sound abstract operators.
\begin{definition}[Concretization]
  A {\em concretization function} \( \gamma: \adom \rightarrow \cdom
  \) defines the meaning of \( \adom \) in terms of a concrete domain
  \( \cdom = \partsof{\cset} \) for some set of concrete objects \(
  \cset \).  An abstract inclusion $\ax_1 \sqsubseteq \ax_2$ for
  abstract elements $\ax_1, \ax_2 \in \adom$ should be sound with
  respect to concrete inclusion: $\gamma(\ax_1) \subseteq
  \gamma(\ax_2)$, and $\gamma$ should be monotone.  For each concrete
  operation $f$,  we expect a sound abstract counterpart $\af$; for example,
  an abstract operation \( \af: \adom \rightarrow \adom \) is sound
  with respect to a concrete operation \( f: \cdom \rightarrow \cdom \)
  if and only if $\gamma( \ax ) \subseteq \gamma \circ \af( \ax )$ for
  all $\ax \in \adom$.
\end{definition}

In this section, we focus on the abstract domains and concretization
functions, while the construction of abstract operations are
detailed in Section~\ref{sec:4:ops}.

\subsection{An exact store abstraction based on separating shape graphs}
\label{sec:3:1:ssg}

%
An {\em abstract heap} \( \aheap \in \aheaps \) should
over-approximate a set of concrete heaps with a compact
representation. This set of abstract heaps $\aheaps$ form the {\em
  domain of abstract heaps} (or the {\em shape abstract domain}).  For
simplicity, we first consider an \emph{exact abstraction} of heaps
with no unbounded dynamic data structures.  That is, such an
abstraction explicitly enumerates a finite number of memory cells and
performs no summarization.  Summarization is considered in
Section~\ref{sec:3:3:summarize}.

%
A heap can be viewed as a set of {\em disjoint} cells (cf.,
Figure~\ref{fig:1:exmem}).  At the abstract level, it is convenient to
make disjointness explicit and describe disjoint cells independently.
Thus, we write \( \aheap_0 \lsep \aheap_1 \) for the abstract heap
element that denotes all that can be partitioned into a sub-heap
satisfying \( \aheap_0 \) and another disjoint sub-heap satisfying \(
\aheap_1 \).  This observation about disjointness underlies separation
logic~\cite{r:lics:02} and thus we borrow the separating conjunction
operator \( \lsep \) from there.
An individual cell is described by an {\em exact points-to} predicate
of the form \( \alpha \cdot \fldf \pt \beta \) where \( \alpha, \beta
\) are symbolic variables (or, abstract values) drawn from a set \(
\avars \).  The symbolic variable $\alpha$ denotes an address, while
$\beta$ represents the contents at the memory cell with address
$\alpha \cdot \fldf$ (i.e., $\alpha$ offset by a field $\fldf$).  An
exact heap abstraction is thus a separating conjunction of a set of
exact points-to predicates.

\begin{wrapfigure}{r}{0pt}\centering
  \includegraphics{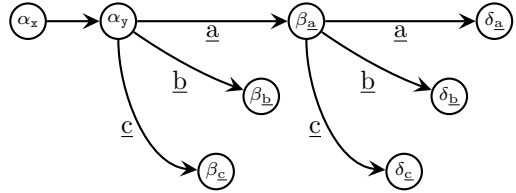}
  \caption{ separating shape graph abstraction of $\heap$ in
    Figure~\ref{fig:1:exmem}.  Symbolics $\alpha_{\varx}$
    and \( \alpha_{\vary} \) denote the {\em address} of
    $\varx$ and \( \vary \), respectively.
}
  \label{fig:2:exssg}
\end{wrapfigure}
Such abstract heap predicates can be represented using {\em separating
  shape graphs}~\cite{xisa:sas:07,xisa:esop:10} where nodes are
symbolic variables and edges represent heap predicates.  An exact
points-to predicate \( \alpha \cdot \fldf \pt \beta \) is denoted by
an edge from node \( \alpha \) to node \( \beta \) with a label for
the field offset $\fldf$.  For example,
\( \beta_{\flda} \) denotes the {\em value} corresponding to the C
expression $\vary \cdot \flda$.


The concretization \( \gammaheap \) of a separating shape graph must
account for symbolic variables that denote some concrete values, so it
also must yield an {\em instantiation} or a {\em valuation} \( \valua:
\avars \rightarrow \svals \).  Thus, this concretization has type
$\gammaheap : \aheaps \rightarrow \partsof{\sheaps \times (\avars
  \rightarrow \svals)}$ and is defined as follows (by induction on the
structure $\aheap$):
\[ \begin{array}{@{}r@{\;}c@{\;}l@{}}
\gammaheap( \alpha \cdot \fldf \pt \beta )
  & \defeq &
  \{
  ([\valua(\alpha) + \fldf \mapsto \valua(\beta)], \valua)
  \mid
  \valua \in \avars \rightarrow \svals
  \}
\\
\gammaheap( \aheap_0 \lsep \aheap_1 )
  & \defeq &
  \{
  (\heap_0 \uplus \heap_1, \valua)
  \mid
  (\heap_0, \valua) \in \gammaheap( \aheap_0 )
  \;\text{and}\;
  (\heap_1, \valua) \in \gammaheap( \aheap_1 )
  \;\text{and}\;
  \heapdom(\heap_0) \cap \heapdom(\heap_1) = \emptyset
  \}
  \;.
\end{array}
\]
That is, an exact points-to predicate corresponds to a single cell
concrete store under a valuation $\valua$, and a separating
conjunction of abstract heaps is a concrete store composed of disjoint
sub-stores that are individually abstracted by the conjuncts under the
same instantiation (as in separation logic~\cite{r:lics:02}).
Symbolic variables can be viewed as existentially-quantified variables
that are bound at the top-level of the abstraction.  The valuation
makes this explicit and thus is a bit similar to a concrete
environment $\env$.


\relatedworkpara Separating conjunction manifests itself in separating
shape graphs as simply distinct edges.  In other words, distinct edges
denote disjoint heap regions.  Separating shape graphs are visually
quite similar to classical shape and points-to
graphs~\cite{chase:pldi:90,tvla:toplas:02} but are
actually quite different semantically.  In classical shape and
points-to graphs, the nodes represent memory cells, and typically, a
node corresponds to one-or-more concrete cells.  Distinct nodes
represent disjoint memory memory regions, and edges express variants
of may or must points-to relations between two sets of cells.  In
contrast, it is the edges in separating shape graphs that correspond
to disjoint memory cells, while the nodes simply represent values.  We
have found two main advantages of this approach.  First, because there
is no \emph{a priori} requirement that two nodes be distinct values,
we do not need to case split simply to speak about the contents of
cells (e.g., consider two pointer variables $\varx$ and $\vary$ and
representing to which objects they point; a classic shape graph must
consider two cases where $\varx$ and $\vary$ are aliases or not, while
a separating shape graph does not).  Limiting case splits is critical
to getting good analysis performance~\cite{bec:phd:2008}.  Second, a
separating shape graph is agnostic to the type of values that nodes
represent.  Nodes may represent addresses, but they can just as easily
represent non-address values, such as integer, Boolean, or
floating-point values.  We take advantage of this observation to
interface with numeric abstract domains~\cite{xisa:popl:08}, which we
discuss further next in Section~\ref{sec:3:2:combin}.


\subsection{Enriching shapes with a numeric abstraction}
\label{sec:3:2:combin}

\begin{figure}[t]
  \begin{center}
    \includegraphics[scale=1]{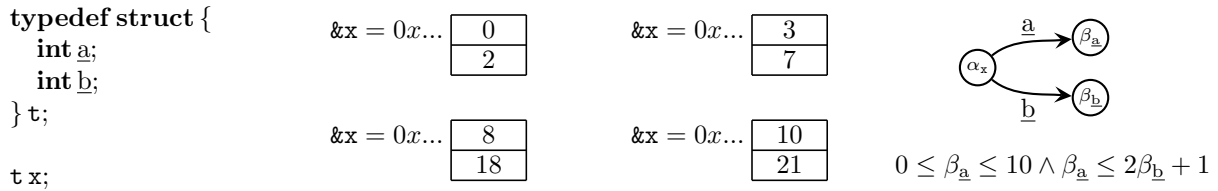}
  \end{center}
  \caption{An example separating shape graph enriched with a numeric
    constraint (right) with four concrete instances (center) for the
    C type declaration (left).}
  \label{fig:3:val}
\end{figure}

From Section~\ref{sec:3:1:ssg}, we have an exact heap abstraction
based on a separating shape graph with a finite number of exact
points-to edges.  Intuitively, this abstraction is quite weak, as we
have simply enumerated the memory cells of interest.  We have,
however, given names to all values---both addresses and contents---of
potential interest.
Here, we enrich the abstraction with information about the values
contained in data structures, not just the pointer shape.  We focus on
{\em scalar} numeric values, such as integers or floating-point
values, but other types of values could be handled similarly.  A
separating shape graph defines a set of symbolic variables
corresponding to values, so we can abstract the values those symbolic
variables represent.
%
First, we consider a simple example, shown in Figure~\ref{fig:3:val}.
In Figure~\ref{fig:3:val}, we show four concrete stores such that \( 0
\leq \varx \cdot \flda \leq 10 \) and \( \varx \cdot \flda \leq
2(\varx \cdot \fldb) + 1 \).  The separating shape graph on the right
clearly abstracts the shape of the four stores (i.e., two fields
$\flda$ and $\fldb$ off a $\cstruct$ at variable $\varx$).  The
symbolic variables $\beta_{\flda}$ and $\beta_{\fldb}$ represent the
contents of cells $\var \cdot \flda$ and $\var \cdot \fldb$,
respectively, so the numeric property specified above can expressed
simply by using a logical formula involving $\beta_{\flda}$ and
$\beta_{\fldb}$ (as shown).

In general, a separating shape graph \( \aheap \) is defined over a
set of symbolic variables \( \gvars{\aheap} \) where \( \gvars{\aheap}
\subseteq \avars \).  The properties of the values stored in heaps
described by \( \aheap \) can be characterized by logical formulas
over \( \gvars{\aheap} \).  Such logical formulas expressing numeric
properties can be represented using a numeric abstract domain \(
\adomnump{\gvars{\aheap}} \) that abstracts functions from \(
\gvars{\aheap} \) to \( \svals \), that is, it comes with
concretization function parametrized by a set of symbolic values
$\gvars{\aheap}$ of the following type: \( \gammanump{\gvars{\aheap}}:
\adomnump{\gvars{\aheap}} \rightarrow \partsof{\gvars{\aheap}
  \rightarrow \svals} \).  For example, the numeric property mentioned
in Figure~\ref{fig:3:val} could be expressed using the convex polyhedra
abstract domain~\cite{ch:popl:78}.  As a shape graph concretizes into
a set of pairs composed of a heap \( \heap \) and a valuation \(
\valua: \gvars{\aheap} \rightarrow \svals \), such numeric constraints
simply restrict the set of admissible valuations.

The need to combine a shape graph with a numeric constraint suggests
using a product abstraction~\cite{cc:popl:79} of a shape abstract
domain \( \aheaps \) and a numeric abstract domain \( \adomnump{-} \).
However, note that the numeric abstract domain that needs to be used
depends on the separating shape graph, as the set of dimensions is
equal to the set of nodes in the separating shape graph.  Therefore,
the conventional notion of a {\em symmetric} reduced product does not
apply here.  Instead, we use a different construction known as a {\em
  cofibered abstract domain}~\cite{venet:sas:96} (in reference with
the categorical notion underlying this construction). 
\begin{definition}[Combined shape-numeric abstract domain]
  Given a shape domain $\aheaps$ and a numeric domain $\adomnump{-}$
  parametrized by a set of symbolic variables.  We let \( \adomnum \)
  denote the set of numeric abstract values corresponding to any shape
  graph (i.e., \( \adomnum \defeq \bigcup \{ \adomnump{V} \mid V
  \subseteq \avars \} \)), and we define the {\em combined
    shape-numeric abstract domain} \( \adomcomb \) and its
  concretization \( \gammacomb: (\adomcomb)
  \rightarrow \partsof{\sheaps \times (\avars \rightarrow \svals)} \)
  as follows:
\[
\begin{array}{rcl}
  \adomcomb
  & \defeq
  & \{ (\aheap, \anum) \mid \aheap \in \aheaps
  \;\text{and}\;
  \anum \in \adomnump{\gvars{\aheap}} \}
  \\
  \gammacomb( \aheap, \anum )
  & \defeq
  & \{ (\heap, \valua)
  \mid
  (\heap, \valua) \in \gammaheap( \aheap )
  \;\text{and}\;
  \valua \in \gammanump{\gvars{\aheap}}( \anum ) \}
  \\
\end{array}
\]
\end{definition}
\begin{figure}[t]
  \begin{center}
    \includegraphics[scale=1]{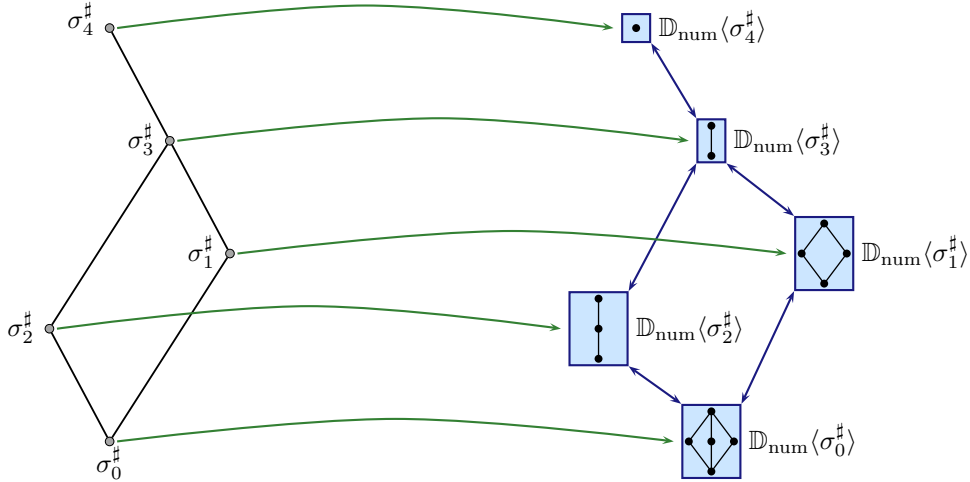}
  \end{center}
  \caption{The combined shape-numeric abstract domain is a cofibered
    layering of a numeric abstract domain on a shape abstract domain.}
  \label{fig:5:cofib}
\end{figure}
This product is clearly {\em asymmetric}, as the left member defines
the abstract lattice to which the right member belongs.  We illustrate
this structure in Figure~\ref{fig:5:cofib}.  The left part depicts the
lattice of abstract heaps, while the right part illustrates a lattice
of numeric lattices.  Each element of the lattice of lattices is an
instance of the numeric abstract domain over the symbolic variables
defined by the abstract heap, that is, it is the image of the function
\( \aheap \mapsto \adomnump{\gvars{\aheap}} \).


\begin{figure}[t]\centering
    \includegraphics[scale=1]{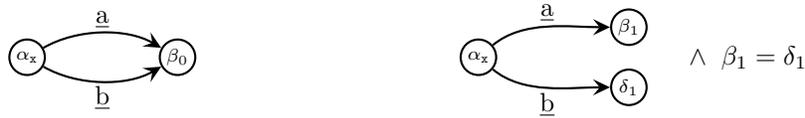}
  \caption{Two abstractions drawn from the combined abstract domain
    $\adomcomb$ that have equivalent concretizations but with
    non-isomorphic sets of symbolic variables.}
  \label{fig:4:asym}
\end{figure}
\begin{figure}\centering
  \includegraphics[scale=1]{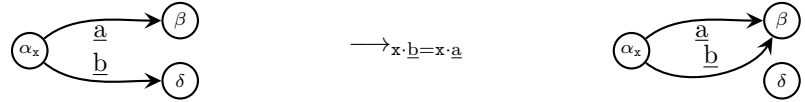}
  \caption{Applying the transfer function for an assignment on a
    separating shape graph that changes the set of ``live'' symbolic
    variables.}
  \label{fig:4:asym-transfer}
\end{figure}

This dependence is not simply theoretical but has practical
implications on both the representation of abstract values and the
design of abstract operations in the combined abstract domain.
For instance, Figure~\ref{fig:4:asym} shows two separating shape graphs
together with numerical invariants that represent the same set of concrete
stores even though they use two different sets of symbolic variables (even up
to $\alpha$-renaming).  Both of these combined shape-numeric abstract
domain elements represent a store with two fields $\varx \cdot \flda$
and $\varx \cdot \fldb$ such that $\varx \cdot \flda = \varx \cdot
\fldb$.
In the right abstract domain element, the contents of both fields are
associated with distinct nodes, and the values denoted by those nodes
are constrained to be equal by the numeric domain. In the left
graph, the contents of both fields are associated to the same node,
which implies that they must be equal (without any constraint in the
numeric domain).

Now, with respect to the design of abstract operations in the combined
abstract domain, the set of nodes in the shape graph will in general
change during the course of the analysis.  For instance, the analysis
of an assignment of the value contained into field \( \flda \) to
field \( \fldb \) from the abstract state shown in the left produces
the one in the right in Figure~\ref{fig:4:asym-transfer}.  After this
transformation takes place, node \( \delta \) becomes ``garbage'' or
irrelevant, as it is not linked anywhere in the shape graph, and no
numeric property is attached to it.  This symbolic variable $\delta$
should thus be removed or projected from the numeric abstract domain.
Other operations can cause new symbolic variables to be added, and
this issue is only magnified with summaries (cf.,
Section~\ref{sec:3:3:summarize}).  Thus, the combined abstract domain
must take great care in ensuring the consistency of the numeric
abstract values with the shape graphs, as well as dealing with graphs
with different sets of nodes.  Considering again the diagram in
Figure~\ref{fig:5:cofib}, whenever two shape graphs are ordered
$\aheap_0 \sqsubseteq \aheap_1$, there exists a \emph{symbolic
  variable renaming function} \( \symconv{\aheap_0,\aheap_1}:
\gvars{\aheap_1} \rightarrow \gvars{\aheap_0} \) that expresses a
renaming of the symbolic variables from the weaker shape graph
$\aheap_1$ to the stronger one $\aheap_0$.  For example, the symbolic
renaming function $\symconvsym$ for the shape graphs shown in
Figure~\ref{fig:4:asym} is $[\alpha_{\varx} \mapsto \alpha_{\varx},
\beta_1 \mapsto \beta_0, \delta_1 \mapsto \beta_0]$.



%

\relatedworkpara
In practice, the implementation of the shape abstract domain takes the form
of a functor (in the ML programming sense) that takes as input a module
implementing a numeric domain interface (\eg, a wrapper on top of the
\apron library~\cite{apron:cav:09}) and outputs another module that
implements the memory abstract domain interface.
The construction that we have shown in this section is general to analyses
where the set of symbolic variables is dynamic during the course of
the analysis and where the inference
of this set is bound to the inference of cell contents.
In other words, it is well-suited to applying shape analyses for
summarizing memory cells
and then reasoning about their contents with another domain.
This construction has been used not
only in Xisa~\cite{xisa:popl:08} but also in a TVLA-based
setup~\cite{bill:sas:10} and one based on a history of heap
updates~\cite{bec:vmcai:05}.

Another approach that avoids this construction by performing a
sequence of analyses: first, a shape analysis infers the set of
symbolic variables; then, a numeric static analysis relies on this
set~\cite{magill:sas:07,magill:popl:10}.  While less
involved, this approach prevents the exchange of information between
both analyses, which is often required to achieve a satisfactory
level of precision~\cite{xisa:popl:08}.  This sequencing of heap
analysis followed by value analysis is similar to the application of a
pre-pass pointer analysis followed by model checking over a Boolean
abstraction exemplified in SLAM~\cite{ball:pldi:01} and
BLAST~\cite{henzinger:popl:02}

\subsection{Enhancing store abstractions with summaries}
\label{sec:3:3:summarize}

So far, we have considered very simple abstract heaps described by
separating shape graphs where all concrete memory cells are abstracted
by exact points-to edges.  To support abstracting a potentially
unbounded number of concrete memory cells via dynamic memory
allocation, we must extend abstract heaps with {\em summarization},
that is, a way of providing a compact abstraction for possibly
unbounded, possibly non-contiguous memory regions.

\begin{wrapfigure}{r}{0pt}
  \includegraphics[scale=1]{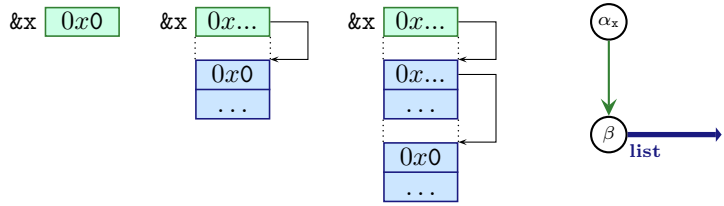}
  \caption{Summarizing linked lists with inductive predicate edges in
    separating shape graphs.}
  \label{fig::exsum}
\end{wrapfigure}
As an example, consider the concrete stores shown in the left part of
Figure~\ref{fig::exsum} consisting of a series of linked lists with 0,
1, and 2 elements. These concrete stores are just instances among
infinitely many ones where \( \varx \) stores a reference to a list of
arbitrary length.  Each of these instances consist of two regions: the
cell corresponding to variable \( \varx \) (green) and the list
elements (blue).  To abstract all of these stores in a compact and
precise manner, we need to summarize the second region with a
predicate.  We can define such a predicate for summarizing such a
region using an inductive
definition \( \indlist \) following the structure of lists: \(
\icallpz{\alpha}{\indlist}:= (\emp \wedge \alpha = 0x\mathtt{0}) \vee
(\alpha \cdot \flda \pt \beta_0 \lsep \alpha \cdot \fldb \pt \beta_1
\lsep \icallpz{\beta_0}{\indlist} \wedge \alpha \not= 0x\mathtt{0})
\).  This definition notation is slightly non-standard to match the graphical
notation: the predicate name is
$\indlist$ and $\alpha$ is the formal induction parameter.
A $\indlist$ memory region is empty if the root pointer
$\alpha$ of the list is null, or
otherwise, there is a head list element with two fields $\flda$ and
$\fldb$ such that the contents of cell $\alpha \cdot \flda$ called
$\beta_0$ is itself a pointer to a list.  Then, in
Figure~\ref{fig::exsum}, if variable \( \varx \) contains a pointer
value denoted by \( \beta \), the second region can be summarized by
the inductive predicate instance \( \icallpz{ \beta}{\indlist} \).
Furthermore, the three concrete stores are abstracted by the abstract
heap \( \alpha_{\varx} \pt \beta \lsep \icallpz{\beta}{\indlist} \) (drawn as
a graph to the right).  The inductive predicate \( \icallpz{\beta}{\indlist} \)
is drawn as the bold, thick edge from node \( \beta \).

\paragraph{Materialization:}

The analyzer must be able to apply transfer functions on summarized
regions.  However, designing precise transfer functions on arbitrary
summaries is extremely difficult.  An effective approach is to define
direct transfer functions only on exact predicates and then define
transfer functions on summaries indirectly via
\emph{materialization}~\cite{sagiv:toplas:98} of exact predicates from
them.
In the following, we focus on the case where summaries are derived
from inductive predicates~\cite{xisa:sas:07} and thus call the
materialization operation {\em unfolding}.  In practice, unfolding
should be guided by a specification of the summarized region where the
analyzer needs to perform local reasoning on materialized cells (see
Section~\ref{sec:4:2:unfold}).  However, from the theoretical point of
view, we can let an unfolding operator be defined as some function that
replaces one abstract \( (\aheap, \anum) \) with a {\em finite set} of
abstract elements \( (\aheap_0, \anum_0), \ldots, (\aheap_{n-1},
\anum_{n-1}) \).
\begin{definition}[Materialization]
  Let us write \( \relunfold \subseteq (\adomcomb)
  \times \partsoffin{\adomcomb} \) for the unfolding relation.  Then,
  any unfolding of an abstract element should be sound with respect
  to concretization:
\[
\text{If}
\;
  (\aheap, \anum) \relunfold
  (\aheap_0, \anum_0), \ldots, (\aheap_{n-1}, \anum_{n-1})
\;,
\text{then}
\;
\gammacomb( \aheap, \anum ) \subseteq
\bigcup_{0 \leq i < n} \gammacomb( \aheap_{i}, \anum_{i})
\;.
\]
As seen above, the finite set of abstract elements that results from
materialization represents a disjunction of abstract elements (i.e.,
materialization is a form of case analysis).
For precision, we typically want an equality instead of
inclusion in the conclusion, which motivates a need to represent a
disjunction of abstract elements (cf., Section~\ref{sec:3:4:env}).
\end{definition}
\begin{example}[Unfolding an inductively-defined list]
  For instance, the abstract element from $\adomcomb$ depicted in
  Figure~\ref{fig::exsum} can be unfolded to two elements:
  \[
  (\alpha_{\varx} \pt \beta \lsep \icallpz{\beta}{\indlist}, \top)
  \relunfold
  (\alpha_{\varx} \pt \beta, \beta = 0x\mathtt{0}), (\alpha_{\varx}
  \pt \beta \lsep \beta \cdot \flda \pt \beta_0 \lsep \beta \cdot
  \fldb \pt \beta_1 \lsep \icallpz{\beta_0}{\indlist}, \beta \not=
  0x\mathtt{0})
  \;,
  \]
  which means that the list pointer $\beta$ is either a null pointer
  or points to a list element whose $\flda$ field contains a pointer
  to another list.
\end{example}

\newcommand{\predls}{\mathtt{ls}}
\relatedworkpara
Historically, the idea of using compact summaries for an unbounded
number of concrete memory cells goes back to at least Jones and
Muchnick~\cite{jones+1981:flow-analysis}, though the set of abstract
locations was fixed \emph{a priori} before the analysis.  Chase et
al.~\cite{chase:pldi:90} considered dynamic
summarization during analysis, while Sagiv et al.~\cite{sagiv:toplas:98}
introduced materialization.
We make note of existing analysis algorithms that make use of
summarization-materialization. \emph{TVLA summary
  nodes}~\cite{tvla:toplas:02} represent unbounded sets of concrete
memory cells with predicates that express universal properties of all
the concrete cells they denote.  The use of three-valued logic enables
abstraction beyond a set of exact points-to constraints (i.e., the
separating shape graphs in Section~\ref{sec:3:1:ssg} are akin to
two-valued structures in TVLA), and summarization is controlled by
instrumentation predicates that limits the compaction done by
canonical abstraction.
Fixed {\em list segment predicates}~\cite{dino:tacas:06,berdine:cav:07}
  characterize consecutive chains of list elements by its first and last
  pointers.  Thus, a predicate of the form \( \predls( \alpha, \alpha') \)
  denotes all chains of list elements (of any length) starting at
  \( \alpha \) and ending at \( \alpha' \).
  Then, an abstract heap consists of a separating conjunction of points-to
  predicates (Section~\ref{sec:3:1:ssg}) and list segments.  These
  predicates can be generalized to other structure segments.
{\em Inductive predicates}~\cite{xisa:sas:07,xisa:popl:08}
  generalize the list segment predicates in several ways.
  First, the abstract domain may be parametrized by a set of
  user-supplied inductive
  definitions.
  Note that as parameters to the abstract domain and thus the
  analyzer, the inductive definitions specify possible templates for
  summarization.  A sound analysis can only infer a summary predicate
  essentially if it exhibits an exact instance of the summary.  The
  ``correctness'' of such inductive definitions are not assumed, but
  rather a disconnect between the user's intent and the meaning
  an inductive predicate could lead to unexpected results.
 Second, inductive predicates can correspond to
  complete structures (\eg, a tree that is completely
  summarized into a single abstract predicate), whereas segments
  correspond to incomplete structures characterized by a
  missing sub-structure.  Inductive predicates can be generically
  lifted to unmaterializable segment summaries~\cite{xisa:sas:07} or
  materializable ones~\cite{xisa:popl:08}.
Independently, {\em array region predicates}~\cite{gopan:popl:05} have been used to
  describe the contents of zones in arrays.  Some analyses on arrays and
  containers have used index variables into summaries instead of
  explicit materialization
  operations~\cite{peron:pldi:08,gulwani:popl:08,dillig:popl:11}.

\subsection{Lifting store abstractions to disjunctive memory state abstractions}
\label{sec:3:4:env}

At this point, we have described an abstraction framework for concrete
stores $\heap$.  To complete an abstraction for memory states $\mem:
(\env, \heap)$, we need two things: (1) an abstract counterpart to
$\env$ and (2) a disjunctive abstraction for when a single abstract
heap $\aheap$ is insufficient for precisely abstracting the set of
possible concrete stores.

\paragraph{Abstract environments:}
Since the abstract counterpart for addresses are
symbolic variables (or nodes) in shape graphs, an \emph{abstract
  environment} \( \aenv \) can simply be a function mapping program
variables to nodes, that is, \( \aenv \in \aenvs = \svars \rightarrow
\avars \).
Now, the {\em memory abstract domain} \( \adommem \) is defined by \( \adommem =
\aenvs \times (\adomcomb) \), and its concretization \( \gammamem :
\adommem \rightarrow \partsof{\senvs \times \sheaps} \)
can be defined as follows:
\[
\gammamem(\aenv, (\aheap, \anum)) \defeq
\{ (\valua \circ \aenv, \heap) \mid
(\heap, \valua) \in \gammaheap( \aheap )
\;\text{and}\;
\valua \in \gammanump{\gvars{\aheap}}( \anum )
\}
\;.
\]
Note that in an abstract memory state $\amem: (\aenv, \aheap)$, the
abstract environment $\aenv$ simply gives the symbolic address of
program variables, while the abstract heap $\aheap$ abstracts all
memory cells---just like the concrete model in
Section~\ref{sec:2:1:concrete-semantics}.

\begin{wrapfigure}{r}{0pt}
  \includegraphics[scale=1]{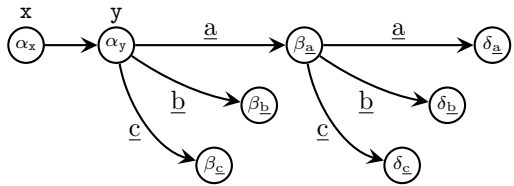}
  \caption{Depicting a memory abstraction including the abstract heap
    from Figure~\ref{fig:2:exssg} and an abstract environment.}
  \label{fig::stateabs}
\end{wrapfigure}
We let the abstract environment be depicted by node labels in the
graphical representation of abstract heaps.  For instance, the
concrete memory state shown in Figure~\ref{fig:1:exmem} can be
described by the diagram in Figure~\ref{fig::stateabs}.

\paragraph{Disjunctive abstraction:}

Recall that the unfolding operation from
Section~\ref{sec:3:3:summarize}
generates a finite disjunction of abstract facts---specifically,
combined shape-numeric abstract elements $\{ \ldots, (\aheap_i,
\anum_i), \ldots \} \subseteq \adomcomb$.
Thus, a disjunctive abstraction layer is required regardless of other analysis reasons
(e.g., path-sensitivity).
We assume the \emph{disjunctive abstraction} is defined by an abstract
domain \( \adomdisj \) and a concretization function \( \gammadisj:
\adomdisj \rightarrow \partsof{\smems} \).  We do not prescribe any
specific disjunctive abstraction.  A simple choice is to apply a
disjunctive completion~\cite{cc:popl:79}, but further innovations
might be possible by taking advantage of being specific to memory.
\begin{example}[Disjunctive completion]
For a memory abstract domain $\adommem$, its disjunctive completion
$\adomdisj$ is defined as follows:
\[
\adomdisj \defeq \partsoffin{\adommem}
\qquad\qquad
\gammadisj( \astate ) \defeq \bigcup \{ \gammamem( \amem ) \mid \amem
\in \astate \}
\;.
\]
\end{example}

\begin{figure}[tb]\centering
  \includegraphics[scale=1]{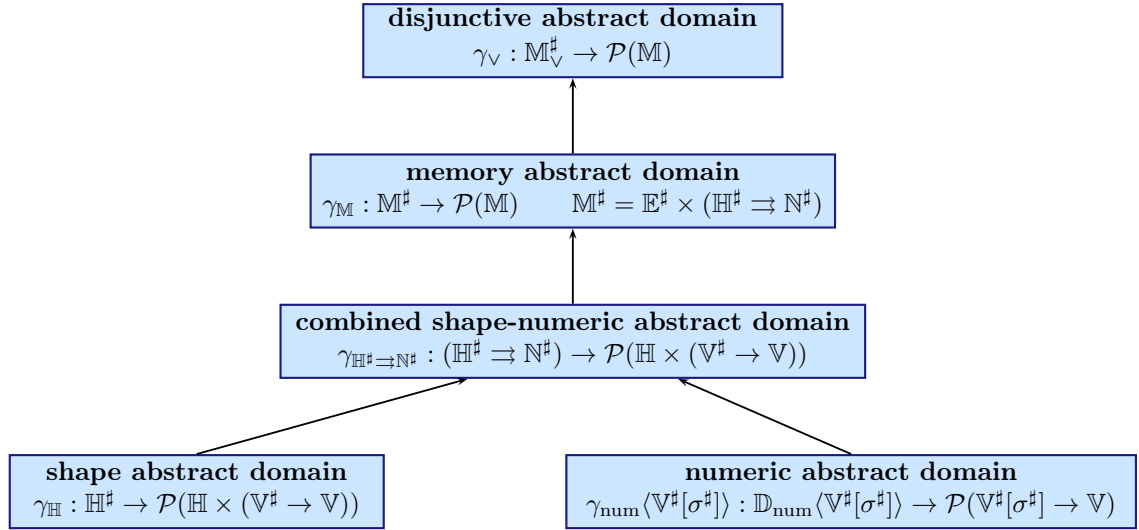}
  \caption{Layers of abstract domains to yield a disjunctive memory
    state abstraction.  From an implementation perspective, the edges
    correspond to inputs for ML-style functor instantiations.}
  \label{fig::domstruct}
\end{figure}
In Figure~\ref{fig::domstruct}, we sum up the structure of the
abstract domain for abstracting memory states $\smems$
as a stack of layers, which are typically implemented as ML-style functors.
Each layer corresponds to the abstraction of a different form of concrete
semantics (as shown in the diagram).


\relatedworkpara Trace partitioning~\cite{rm:toplas:07} relies on
control-flow history to manage disjunctions, which could be used as an
alternative to disjunctive completion.  However, it is a rather general
construction and can be instantiated in multiple ways with a large
effect on precision and performance.

\begin{figure}[p]
  \begin{compactitem}
  \item A shape abstract domain $\aheaps$
    \[
    \begin{array}{rlcl}
      \evallheap:
      & \lvals{\avars} \times \aheaps
      & \longrightarrow
      & \undefset{(\avars \times \sfields \times \aheaps)}
      \\
      \evaleheap:
      & \exprs{\avars} \times \aheaps
      & \longrightarrow
      & \undefset{(\avars \times \aheaps)}
      \\
      \mutateheap:
      & \avars \times \sfields \times \avars \times \aheaps
      & \longrightarrow
      & \undefset{\aheaps}
      \\
      \unfoldheap:
      & (\lvals{\avars} \times \sfields) \times \aheaps
      & \longrightarrow
      & \partsoffin{\aheaps \times \exprs{\avars}}
      \\
      \newheap:
      & \avars \times \aheaps
      & \longrightarrow
      & \aheaps
      \\
      \deletenheap:
      & \avars \times \aheaps
      & \longrightarrow
      & \aheaps
      \\
      \deleteeheap:
      & \avars \times \sfields \times \aheaps
      & \longrightarrow
      & \aheaps
      \\
      \compareheap:
      & (\avars \rightarrow \avars) \times \aheaps \times \aheaps
      & \longrightarrow
      & \{ \false \} \uplus \{ \true \} \times (\avars \rightarrow \avars)
      \\
      \joinheap:
      & ((\avars)^2 \rightarrow \avars) \times \aheaps \times \aheaps
      & \longrightarrow
      & \aheaps \times (\avars \rightarrow \avars)^2
      \\
      \widenheap:
      & ((\avars)^2 \rightarrow \avars) \times \aheaps \times \aheaps
      & \longrightarrow
      & \aheaps \times (\avars \rightarrow \avars)^2
      \\
    \end{array}
    \]
  \item A numeric abstract domain over symbolic variables $\adomnum$
    \[
    \begin{array}{rlcl}
      \assignnum:
      & \lvals{\avars} \times \exprs{\avars} \times \adomnum
      & \longrightarrow
      & \adomnum
      \\
      \guardnum:
      & \exprs{\avars} \times \adomnum
      & \longrightarrow
      & \adomnum
      \\
      \newnum:
      & \avars \times \adomnum
      & \longrightarrow
      & \adomnum
      \\
      \deletennum:
      & \avars \times \adomnum
      & \longrightarrow
      & \adomnum
      \\
      \renamenum:
      & (\avars \rightarrow \avars) \times \adomnum
      & \longrightarrow
      & \adomnum
      \\
      \comparenum:
      & \adomnum \times \adomnum
      & \longrightarrow
      & \sbools
      \\
      \joinnum:
      & \adomnum \times \adomnum
      & \longrightarrow
      & \adomnum
      \\
      \widennum:
      & \adomnum \times \adomnum
      & \longrightarrow
      & \adomnum
      \\
    \end{array}
    \]
  \item A combined shape-numeric abstract domain $\adomcomb$
    \[
    \begin{array}{rlcl}
      \assigncomb:
      & \lvals{\avars} \times \exprs{\avars} \times \adomcomb
      & \longrightarrow
      & \undefset{\partsoffin{\adomcomb}}
      \\
      \guardcomb:
      & \exprs{\avars} \times (\adomcomb)
      & \longrightarrow
      & \undefset{\partsoffin{\adomcomb}}
      \\
      \unfoldcomb:
      & \lvals{\avars} \times \adomcomb
      & \longrightarrow
      & \undefset{\partsoffin{\adomcomb}}
      \\
      \alloccomb:
      & \lvals{\avars} \times \sfields^{\ast} \times (\adomcomb)
      & \longrightarrow
      & \undefset{\partsoffin{\adomcomb}}
      \\
      \freecomb:
      & \lvals{\avars} \times \sfields^{\ast} \times (\adomcomb)
      & \longrightarrow
      & \undefset{\partsoffin{\adomcomb}}
      \\
      \comparecomb:
      & (\avars \rightarrow \avars) \times (\adomcomb) \times (\adomcomb)
      & \longrightarrow
      & \{ \false \} \uplus \{ \true \} \times (\avars \rightarrow \avars)
      \\
      \joincomb:
      & ((\avars)^2 \rightarrow \avars) \times (\adomcomb) \times (\adomcomb)
      & \longrightarrow
      & (\adomcomb)
      \\
      \widencomb:
      & ((\avars)^2 \rightarrow \avars) \times (\adomcomb) \times (\adomcomb)
      & \longrightarrow
      & (\adomcomb)
      \\
    \end{array}
    \]
  \item A memory abstract domain $\adommem$
    \[
    \begin{array}{rlcl}
      \assignmem:
      & \lvals{\svars} \times \exprs{\svars} \times \adommem
      & \longrightarrow
      & \undefset{\partsoffin{\adommem}}
      \\
      \guardmem:
      & \exprs{\svars} \times \adommem
      & \longrightarrow
      & \undefset{\partsoffin{\adommem}}
      \\
      \allocmem:
      & \lvals{\svars} \times \sfields^{\ast} \times \adommem
      & \longrightarrow
      & \undefset{\partsoffin{\adommem}}
      \\
      \freemem:
      & \lvals{\svars} \times \sfields^{\ast} \times \adommem
      & \longrightarrow
      & \undefset{\partsoffin{\adommem}}
      \\
      \comparemem:
      & \adommem \times \adommem
      & \longrightarrow
      & \sbools
      \\
      \joinmem:
      & \adommem \times \adommem
      & \longrightarrow
      & \adommem
      \\
      \widenmem:
      & \adommem \times \adommem
      & \longrightarrow
      & \adommem
      \\
    \end{array}
    \]
  \end{compactitem}
  \caption{Interfaces for the abstract domain layers shown in
    Figure~\ref{fig::domstruct} (except the disjunctive abstraction
    layer).}
  \label{fig::ifaces}
\end{figure}

\section{Static analysis operations}
\label{sec:4:ops}
In this section, we describe the main abstract operations on the
memory abstract domain $\amems$ and demonstrate how they are computed
through the composition of abstract domains discussed in
Section~\ref{sec:3:abs}.
Our presentation describes each kind of operation (i.e., transfer
functions for commands like assignment, abstract comparison, and
abstract join) one by one and shows how unfolding and folding
operations are triggered by their application.
The end result of this discussion is a description of how these
domains implement the interfaces shown
in Figure~\ref{fig::ifaces}.
For these interfaces, we let \( \sbools \) denote the set of booleans
\( \{ \true, \false \} \) and \( \undef \) denote an undefined value
for some functions that may fail to produce a result.  We write \(
\undefset{X} \) for \( X \uplus \{ \undef \} \) for any set \( X \)
(i.e., an option type).

\subsection{Assignment over materialized cells}
\label{sec:4:1:assign}
First, we consider the transfer function for assignment.
In this subsection, for simplicity, we focus on the case where {\em none} of the
locations
that appear in either side of the assignment are summarized,
and we defer
the case of transfer functions over summarized graph regions to
Section~\ref{sec:4:2:unfold}.  Because of this simplification, the types
of the abstract operators mentioned will not exactly match those given
in Figure~\ref{fig::ifaces}.  At the same time, this transfer function
captures the essence of the shape-numeric combination.

Recall that \( \lval \in \lvals{\svars} \) and \( \expr \in
\exprs{\svars} \) are location and value expressions, respectively, in
our programming language (cf., Figure~\ref{fig:syntax}).  The transfer
function \( \assignmem : \lvals{\svars} \times \exprs{\svars} \times
\adommem \rightarrow \adommem \) should compute a sound post-condition
for the assignment command \( \lval = \expr \) stated as follows:
\begin{condition}[Soundness of $\assignmem$]\label{cond:assignmem}
  If $(\env, \heap) \in \gammamem(\amem)$, then
\[
  (\env, \subst{\heap}{\lvalsem{\lval}( \env, \heap)}{\exprsem{\expr}( \env, \heap )})
  \in \gammamem( \assignmem( \lval, \expr, \amem ) ) \;.
\]
\end{condition}

\begin{figure}[t]\centering
    \includegraphics[scale=1]{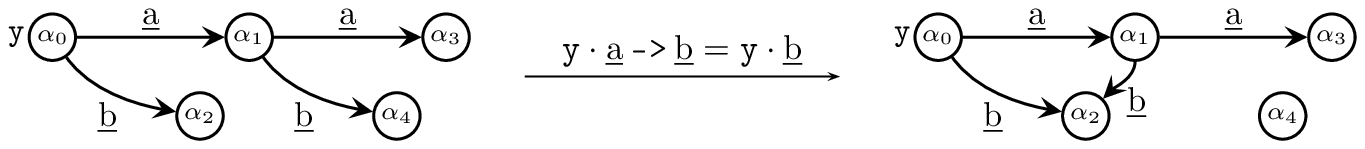}
  \caption{Applying $\assignmem$ to an example assignment of the form
    \( \lval = \lval' \).}
  \label{fig::mutate}
\end{figure}

\paragraph{Assignments of the form \( \lval = \lval' \).}
Let us first assume that right hand side of the assignment is a
location expression.
As an example, consider the assignment shown in
Figure~\ref{fig::mutate} and applying $\assignmem$ to the
pre-condition on the left to yield the post-condition on the right.
The essence is that $\lval$ dictates an edge that should be updated to
point to the node specified by $\lval'$.

To compute a post-condition in this case, \( \assignmem \) should
update the abstract heap, that is, the pre-heap \( \aheap \in \aheaps
\).  An \( \assignmem \) call should eventually forward the assignment
to the heap abstract domain via the $\evallheap$ operation that evaluates a
location expression $\lval$ to an edge, $\evaleheap$ that evaluates
a value expression $\expr$ to a node, and 
$\mutateheap$ that swings a points-to edge.

The base of a sequence of pointer dereferences is given by a program
variable, so the first step consists of replacing the program
variables
in the assignment
with the symbolic names corresponding to their addresses
using the abstract environment $\aenv$.
For our example,
this results in the call to \( \assigncomb( \alpha_0 \cfield \flda
\cdot \fldb, \alpha_0 \cdot \fldb, (\aheap, \anum) ) \) at the
combined shape-numeric layer, which should
satisfy a soundness condition similar to that of \( \assignmem \)
(Condition~\ref{cond:assignmem}).
The next step consists of traversing the abstract heap $\aheap$
according to the location expression and the value expression of the
assignment.
As mentioned above,
this evaluation is performed using
the location evaluation function \( \evallheap
\)
that yields an edge
and the value expression
evaluation function \( \evaleheap
\) that yields a node.
\begin{condition}[Soundness of \( \evallheap \) and \( \evaleheap \)]
Let \( (\heap, \valua) \in \gammaheap( \aheap ) \).  Then,
\[
\begin{array}{l}
\text{If}\;
  \evallheap( \lval, \aheap ) = (\alpha, \fldf)
  \;,
\text{then}\;
  \lvalsem{\lval}( \heap ) = \valua( \alpha ) + \fldf
\;.
\\
\text{If}\;
  \evaleheap( \lval, \aheap ) = \beta
  \;,
\text{then}\;
  \exprsem{\lval}( \heap ) = \valua( \beta )
\;.
\end{array}
\]
\end{condition}

\begin{figure}\centering
\begin{tabular}{@{}p{0.65\linewidth}p{0.25\linewidth}@{}}
\begin{mathpar}
\inferrule[LocAddress]{
}{
  \evallheap( \alpha, \aheap ) = (\alpha, \offzero)
}

\inferrule[LocField]{
  \evallheap( \lval, \aheap ) = (\alpha, \fldf)
}{
  \evallheap( \lval \cdot \fldg, \aheap ) = (\alpha, \fldf + \fldg)
}

\inferrule[ValDereference]{
  \evallheap( \lval, \aheap) = (\alpha, \fldf)
  \\
  \aheap = \aheap_0 \lsep \alpha \cdot \fldf \pt \beta
}{
  \evaleheap( \lval, \aheap) = \beta
}
\end{mathpar}
&
\begin{mathpar}
\inferrule[LocVal]{
  \evaleheap( \expr, \aheap ) = \alpha
}{
  \evallheap( \cderef \expr, \aheap ) = (\alpha , \offzero) 
}

\inferrule[ValLoc]{
  \evallheap( \lval, \aheap ) = (\alpha, \offzero)
}{
  \evaleheap( \caddressof \lval, \aheap ) = \alpha
}
\end{mathpar}
\end{tabular}
\caption{Evaluating dereferences in an abstract heap.}
\label{fig::eval-aheap}
\end{figure}

In Figure~\ref{fig::eval-aheap}, we define $\evallheap$ and
$\evaleheap$ following the syntax of location and value expressions
(over symbolic variables).
We write $\offzero$ for a designated 0-offset
field. This abstract evaluation corresponds
directly to the concrete evaluation defined in
Figure~\ref{fig:opsem-assign}. Note that abstract evaluation is not
necessarily defined for all expressions.  For example, an points-to
edge may simply not exist for the computed address in
\TirName{ValDereference}.  The edge may need to be \emph{materialized} by
unfolding (cf., Section~\ref{sec:4:2:unfold}) or otherwise is a
potential memory error.

Returning to the example in Figure~\ref{fig::mutate}, we get \(
\evallheap( \alpha_0 \cfield \flda \cdot \fldb, \aheap ) = (\alpha_1,
\fldb) \)---the cell being assigned-to corresponds to the exact
points-to edge \( \alpha_1 \cdot \fldb \pt \alpha_4 \)---and \(
\evaleheap( \alpha_0 \cdot \fldb ) = \alpha_2 \)---the value to
assign is abstracted by \( \alpha_2 \).  The abstract post-condition
returned by \( \assigncomb \) should reflect the swinging of that edge in the shape
graph, which is accomplished by the \( \mutateheap \) function:
\[
\mutateheap( \alpha, \fldf, \beta, (\alpha \cdot \fldf \pt \delta) \lsep
\aheap ) = (\alpha \cdot \fldf \pt \beta) \lsep \aheap
\;.
\]
This function simply replaces a points-to edge named by the address
$\alpha$ and field $\fldf$ with a new one for the updated contents
(and fails if such a points-to edge does not exist in the abstract
heap $\aheap$).
The effect of this assignment can be completely reflected in
the abstract heap since the cell corresponding to the assignment is
abstracted by exactly one points-to edge and the new value to store
in that cell is also exactly abstracted by one node.
We note that node \( \alpha_4 \) is no longer reachable in the shape
graph, and thus the value that this node denotes is no longer relevant when
concretizing the abstract state.
As a consequence, it can be safely removed both in \( \aheaps \)
(using function \( \deletenheap \)) and in \( \adomnum \) (using
function \( \deletennum \)).
Such an existential projection or ``garbage collection'' step may be viewed as a conversion operation in
the cofibered lattice structure shown in Figure~\ref{fig:5:cofib}.

\begin{figure}[t]\centering
    \includegraphics[scale=1]{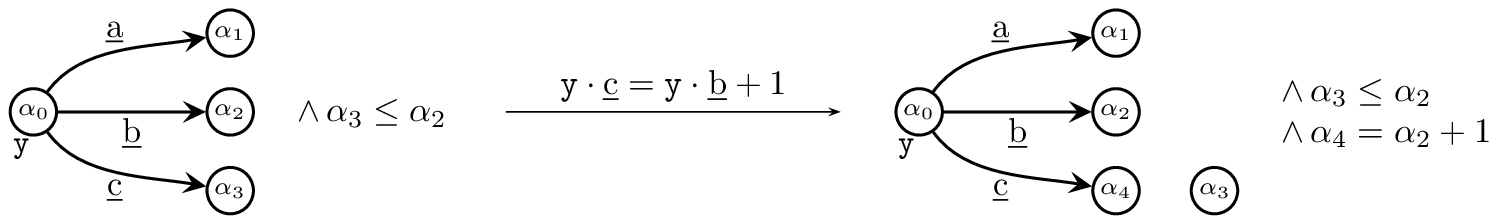}
  \caption{Applying $\assignmem$ to an example assignment of the form
    \( \lval = \expr \).}
  \label{fig::arassign}
\end{figure}

\paragraph{Assignments of the form \( \lval = \expr \).}
In general, the right-hand side of an assignment is not necessarily a
location expression.
The evaluation of left-hand side \( \lval \) proceeds as above,
but the evaluation of the right-hand side expression \( \expr \) is extended.
As an example, consider the assignment shown in
Figure~\ref{fig::arassign}.

The evaluation of the location expression down to the abstract heap
level works as before where we find that \( \evallheap( \alpha_0 \cdot
\fldc, \aheap ) = (\alpha_0, \fldc) \).  For the right-hand--side
expression, it is not obvious what \( \evaleheap( \alpha_0 \cdot \fldb
+ 1, \aheap ) \) should return, as no symbolic node is equal to that
value in the concretization of all elements of \( \aheap \). It is
possible to evaluate sub-expression \( \alpha_0 \cdot \fldb \) to \(
\alpha_2 \), but then \( \evaleheap( \alpha_2 + 1, \aheap ) \) cannot
be evaluated any further.
The solution is to create a new symbolic variable and constrain it
to represent the value of the right-hand--side expression.
Therefore, the evaluation of \( \assigncomb \) proceeds
as follows:
(1) generate a fresh node \( \alpha_4 \);
(2) add \( \alpha_4 \) to the abstract heap \( \aheap \)
and the numeric abstract value \( \anum \) using the function
\( \newheap \) and \( \newnum \), respectively;
(3) update the numeric abstract value \( \anum \) using \( \assignnum(
  \alpha_4, \alpha_2 + 1, \anum ) \), which over-approximates
  constraining $\alpha_4 = \alpha_2 + 1$;
and (4)
mutate with $\mutateheap$ 
with the new node $\alpha_4$ (i.e.,
$\mutateheap(\alpha_0, \fldc, \alpha_4, \aheap)$).

\subsection{Unfolding and assignment over summarized cells}
\label{sec:4:2:unfold}

\begin{wrapfigure}{r}{0pt}\small
\fbox{\begin{math}
\begin{array}{@{}l@{}}
\cstruct \; \clist \, \{ \,
  \cstruct \; \clist \, \cstar \; \fldn; \; \cint \; \fldd;
\}; 
\\[0.5ex]
\begin{array}{@{}r@{\;}r@{\;}l@{}}
\icallpz{\alpha}{\indlist} &:= &(\emp \wedge \alpha = 0x\mathtt{0}) \\
& \vee & (\alpha \cdot \fldn \pt \beta_0 \lsep \alpha \cdot \fldd \pt \beta_1
\lsep \icallpz{\beta_0}{\indlist} \wedge \alpha \not= 0x\mathtt{0})
\end{array}
\end{array}
\end{math}}
\end{wrapfigure}
\begin{figure}[t]\centering
    \includegraphics[scale=1]{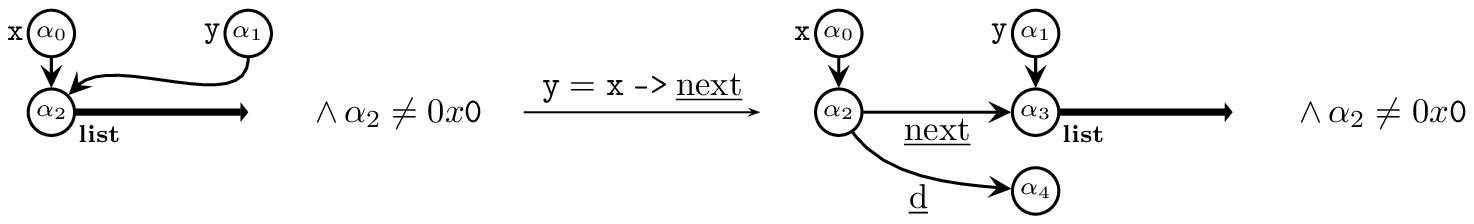}
  \caption{Applying $\assignmem$ to an example that affects the
    summarized region $\alpha_2 \cdot \indlist$.}
  \label{fig::uassign}
\end{figure}

We now consider \( \assignmem \) in the presence of summary
predicates, which intuitively ``get in the way'' of evaluating
location and value expressions in a shape graph.
For instance, consider trying to apply the assignment shown in
Figure~\ref{fig::uassign}.  On the left, we have a separating shape
graph where $\alpha_2$ is a list described by the inductive definition
shown inset.  For clarity, we also show the C-style $\cstruct$
definition that corresponds to the layout of each list element.
%
In applying the assignment, the evaluation of the right-hand--side
expression  \( \varx \cfield \fldn \) fails.  While \( \varx \) evaluates to node
\( \alpha_2 \), there is no points-to edge from \( \alpha_2 \).  Thus,
\( \evaleheap( \alpha_0 \cfield \fldn) \) fails.
It is clear that the reason for this failure is that the memory cell
corresponding to the right-hand--side expression is {\em summarized}
as part of the \( \icallpz{\alpha_2}{\indlist} \) predicate.
To materialize this cell, this predicate should be {\em
  unfolded}; then, the assignment can proceed as in the previous
section (Section~\ref{sec:4:1:assign}).
We can now describe the transfer function for assignment
\( \assignmem( \lval, \expr, (\aheap, \anum) ) \)
in general:
\begin{compactenum}
\item It should call the underlying \( \assigncomb \) and
  follow the process described previously in
  Section~\ref{sec:4:1:assign}.  If evaluation via $\evallheap$ or
  $\evaleheap$ fail, then they should return a failure address, which
  consists of a pair \( (\beta, \fldf) \) corresponding to the node
  and field offset that does not have a materialized points-to edge. In
  the example in Figure~\ref{fig::uassign}, the failure address is \(
  (\alpha_2, \fldn) \).  Note that the interface for evaluation shown
  in Figure~\ref{fig::uassign} does not show the contents of the
  failure case for simplicity.
\item Then, \( \assigncomb \) in the combined domain performs
  an unfolding of the abstract
  heap by calling a function \( \unfoldheap \) that implements
  the unfolding relation \( \relunfold \) with the target points-to edge to materialize
  \( (\beta, \fldf) \).
\begin{condition}[Soundness of $\unfoldheap$]
  \[
    \gammaheap( \aheap )
    \subseteq
    \bigcup \{ (\heap, \valua) \in \gammaheap( \aheap_{\rm u} ) \mid
    (\aheap_{\rm u}, \expr_{\rm u}) \in \unfoldheap( (\beta, \fldf), \aheap )
    \;\text{and}\;
    \sem{\expr_{\rm u}}( \valua ) = \true \}
    \;.
  \]
\end{condition}
Note that unfolding of an abstract heap returns pairs consisting of an
unfolded abstract heap and a numeric constraint as an expression
$\expr_{\rm u} \in \exprs{\gvars{\aheap_{\rm u}}}$ over the symbolic variables
of the unfolded abstract heap.  This expression allows a summary to
contain constraints not expressible in a shape graph itself.  For
instance, in the \( \indlist \) inductive definition, each case comes
with a nullness or non-nullness condition on the head pointer.  Or
more interestingly,
we can imagine an orderedness constraint for an inductive definition
describing an ordered list.  For
the example from Figure~\ref{fig::uassign}, unfolding the shape graph
at $(\alpha_2, \fldn)$ generates two disjuncts, but the one
corresponding to the empty list can be eliminated due to the
constraint that \( \alpha_2 \) has to be non-null.
\item The numeric constraints should be evaluated in the numeric abstract
  domain using a condition test operator $\guardnum$.
  \begin{condition}[Soundness of $\guardnum$]
    Let $V \subseteq \avars$, $\anum \in \adomnump{V}$, and $\valua
    \in \gammanump{V}( \anum )$.  Then,
    \[
    \text{If}\;
    \sem{\expr}( \valua ) = \true
    \;,
    \text{then}\;
    \valua \in \gammanump{V}( \guardnum( \expr, \anum ) )
    \;.
    \]
  \end{condition}
  Thus, the initial abstract state in the combined domain \( (\aheap,
  \anum) \in \adomcomb \) can be over-approximated
  by the following finite set of abstract states:
  \[
  \unfoldcomb(\lval, (\aheap, \anum)) \defeq
  \{ (\aheap_{\rm u}, \guardnum( \expr_{\rm u}, \anum )) \mid
  (\aheap_{\rm u}, \expr_{\rm u}) \in \unfoldheap( (\beta, \fldf), \aheap ) \}
  \]
\item Finally, \( \assigncomb \) should perform the same set of
  operations as described in Section~\ref{sec:4:1:assign} to reflect
  the assignment on {\em each} unfolded heap.  The \( \assigncomb \)
  returns a {\em finite set} of elements because of potential
  unfolding (and similarly for \( \assignmem \)).
  The soundness condition for $\assignmem$ is therefore as
  follows.
  \begin{condition}[Soundness of $\assignmem$]
    Let $(\env, \heap) \in \gammamem( \amem )$.  Then,
  \[
    (\env, \subst{\heap}{\lvalsem{\lval}( \env, \heap)}{\exprsem{\expr}( \env, \heap )})
    \in \bigcup \{ \gammamem( \amem_{\rm u} ) \mid
    \amem_{\rm u} \in \assignmem( \lval, \expr, \amem ) \}
    \;.
  \]
  \end{condition}
  A very similar soundness condition applies to $\assigncomb$.
\end{compactenum}

Figure~\ref{fig::uassign} shows the resulting abstract state for the
assignment after unfolding and mutation on the right.
In certain cases, the unfolding process may have to be performed
multiple times due to repeated failures of calling \( \evallheap \)
and \( \evaleheap \) as shown in Chang and Rival~\cite{xisa:popl:08}.
This behavior is expected, as unfolding may fail to materialize the
correct region, and thus, termination should be enforced with a bound
on the number of unfolding steps.

\subsection{Other transfer functions}
\label{sec:4:3:tfs}
Unfolding is also the basis for most other transfer functions.
Once the points-to edges in question
are materialized, their definition is straightforward as it was for
assignment (cf., Section~\ref{sec:4:1:assign}).
\begin{compactitem}
\item \textbf{Condition test.}
  The abstract domain \( \adommem \) should define an operator \( \guardmem \)
  that takes an expression (of Boolean type) and an abstract value and
  then returns an abstract value that has taken into
  account the effect of the guard expression.
  Just like with assignment, this function may need to perform an unfolding
  and thus returns in general a {\em finite set} of abstract states.
  \begin{condition}[Soundness of $\guardmem$]
    Let $\mem \in \gammamem(\amem)$.  Then,
  \[
  \text{If}\;
    \sem{\expr}( \mem ) = \true
    \;,
    \text{then}\;
    \mem \in \bigcup \{ \gammaheap( \aheap_{\rm u} ) \mid
    \aheap_{\rm u} \in \gammamem( \guardmem( \expr, \amem ) ) \}
    \;.
  \]
  \end{condition}
  It applies the transfer function \( \assignnum \) provided by
  \( \adomnum \) satisfying a similar soundness condition, which is
  fairly standard (\eg, the \apron library provides such a function).
\item \textbf{Memory allocation.}
  Transfer function \( \allocmem \) accounts for the allocation of a
  fresh memory block, and the assignment of the address of this block
  to a given location.
  Given abstract pre-condition \( \aheap \), the abstract allocation
  function \( \allocmem( \lval, [\fldf_1, \ldots, \fldf_n],
  \aheap ) \) returns a sound abstract post-condition for the statement
  \( \lval = \cmalloc(\{\fldf_1, \ldots, \fldf_n\}) \).
\item \textbf{Memory deallocation.}
  Similarly, transfer function \( \freemem \) accounts for freeing
  the block pointed to by an instruction such as \( \cfree \).
  It takes as argument a location pointing to the block being freed,
  a list of fields, and the abstract pre-condition.
  It may also need to perform unfolding to materialize the location.
  It calls \( \freecomb \) in the \( \adomcomb \) level, which then
  materializes points-to edges corresponding to the block to remove
  and deletes them from the graph using function \( \deleteeheap \)
  defined by \( \deleteeheap( \alpha, \fldf, \alpha \cdot \fldf \pt
  \beta \lsep \aheap_0 ) = \aheap_0 \).
  After removing these edges, some symbolic nodes may become
  unreachable in the graph and should be removed using \( \deletenheap
  \) and \( \deletennum \).
\end{compactitem}
The analysis of a more full featured programming language would require
additional classical transfer functions, such as support for variable
creation and deletion, though this can be supported completely at
the memory abstract domain $\adommem$ layer with the abstract
environment $\aenv$.

\begin{figure}[t]\centering
    \includegraphics[scale=1]{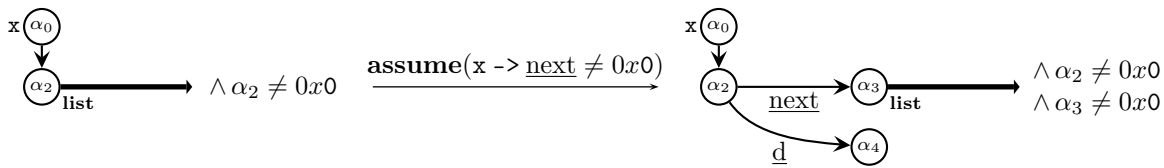}
  \caption{Applying the condition test $\guardmem$ to an example that
    affects a summarized region $\alpha_2 \cdot \indlist$.}
  \label{fig::uguard}
\end{figure}

As an example of a condition test, consider applying $\guardmem$
in Figure~\ref{fig::uguard}.
In the same way as for the example assignment of Figure~\ref{fig::uassign},
the first attempt to compute \( \guardcomb( \alpha_2 \cfield \fldn
\not= 0x\texttt{0}, \aheap ) \) fails, as there is no points-to edge labeled with
\( \fldn \) starting from node \( \alpha_2 \). Thus \( \guardcomb \)
must first call \( \unfoldcomb \).
The unfolding returns a pair of abstract elements, yet the one corresponding
to the case where the list is empty does not need to be considered any
further due to the numerical constraint \( \alpha_2 \not= 0x\texttt{0} \).
Therefore, only the second abstract elements remains, which corresponds
to a list with the first element materialized.
At this stage, expression \( \alpha_2 \cfield \fldn \) can be
evaluated.
Finally, the condition test is reflected by applying \( \guardnum \)
in the numerical
abstract domain \( \adomnum \).

\subsection{Abstract comparison}
\label{sec:4:4:cw}
Abstract interpreters make use of inclusion testing operations in many
situations, such as checking that an abstract post-fixed point has been
reached in a loop invariant computation or that some, for
example,
user-supplied post-condition can be verified with the analysis results.
As inclusion is often not decidable, the comparison function is not
required to be complete but should meet a soundness condition:
\begin{condition}[Soundness of $\comparemem$]
If
$\comparemem( \amem_0, \amem_1 ) = \true$,
then
$\gammamem( \amem_0 ) \subseteq \gammamem( \amem_1)$
\end{condition}
The implementation of such an operator is complicated by
the fact that
the underlying abstract heaps may have {\em distinct} sets of symbolic
nodes.
This issue is a manifestation of the
the cofibered abstract
domain construction (Section~\ref{sec:3:2:combin}).
The concretizations of all abstract domains below \( \adomcomb \)
make use of {\em valuations}, and thus the inclusion checking operator
needs to account for a relation between the symbolic nodes of the
graphs.
This relation between nodes in two graphs \( \relcw \) is computed
step-by-step during the course of the
inclusion checking.

\begin{figure}[t]\centering
    \includegraphics[scale=1]{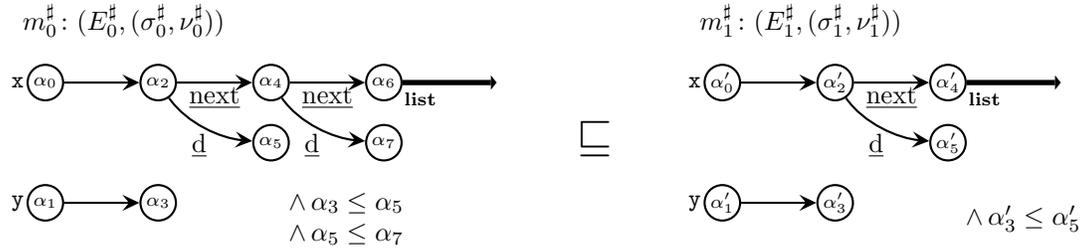}
  \caption{An abstract inclusion that holds and
    shows the need for a node relation $\relcw$.
    In both abstract heaps, variable $\varx$ points to a list and
    $\vary$ points to a number.  On the left, the abstract
    heap describes a list with at least two elements, while on the
    right, it describes one with at least one element.  The
    number pointed to by $\vary$ is less than or equal to
    the data field $\fldd$ of the first element in both abstract
    heaps. The data field of the first element is less than
    or equal to the data field of the second in the left abstract
    heap.}
  \label{fig::cw}
\end{figure}

The example in Figure~\ref{fig::cw} illustrates these difficulties.
It is quite intuitive that any state in the concretization of \( \amem_0 \)
is also in the concretization of \( \amem_1 \).
To see the role of the node relation $\relcw$, let us consider
concretizations of $\amem_0$ and $\amem_1$.
Clearly, if concrete state \( (\env, \heap) \) is in the concretization
of \( \amem_0 \) and in the concretization of $\amem_1$, then node \(
\alpha_0 \) in $\amem_0$ and $\alpha_0'$
denotes the address of \( \varx \).
Thus \( \alpha_0 \)
and \( \alpha'_0 \) denote the {\em same} value, that is, valuations used as
part of the concretization should map those two nodes to the same
value.  The \( \relcw \) should relate these two nodes akin to a
unification substitution.
Similarly, \( \alpha_2 \) and \( \alpha'_2 \) both denote the value
stored in variable \( \varx \), thus should be related in \( \relcw \).
On the other hand, node \( \alpha_6 \) of abstract state \( \amem_0 \) has
no counterpart in \( \amem_1 \)---it corresponds to a null or non-null
address in the region {\em summarized} by the inductive edge.

We notice \( \relcw \) can be viewed as a map from nodes in
\( \amem_1 \) to nodes of \( \amem_0 \) and in this example, defined
by
\( \relcw( \alpha'_{i} ) = \alpha_{i} \) for $0 \leq i \leq 5$.
Also, we notice that mapping \( \relcw \) can be derived
step-by-step, starting from the abstract environments.
Thus, \( \compareheap \) and \( \comparecomb \)
each take as a parameter a set of pairs of symbolic nodes that should be related in
\( \relcw \).  We call this initial set
the {\em roots}, as they are used as a
starting point in the computation of \( \relcw \).

We can now describe the steps of computing
\( \comparemem( \amem_0\colon (\aenv_0, (\aheap_0, \anum_0)),
\amem_1\colon (\aenv_1, (\aheap_1, \anum_1))) \):
\begin{compactenum}
\item First, an initial node mapping $\relcw : \gvars{\aheap_1}
  \rightarrow \gvars{\aheap_0}$
  is derived from the abstract
  environments: \( \relcw \defeq \aenv_0 \circ (\aenv_1)^{-1} \).
  This definition states that the addresses of the program variables in
  $\amem_1$ correspond to the respective addresses of the program
  variables in $\amem_0$.  It is well-defined, as two distinct
  variables cannot be allocated at the same physical address.
\item Then, it calls
  \( \comparecomb( \relcw, (\aheap_0, \anum_0), \) \( (\aheap_1,
  \anum_1)) \) that forwards to a call of
  \( \compareheap( \relcw, \aheap_0, \aheap_1 ) \).
\item The abstract heap comparison function \( \compareheap \) attempts to
  match \( \aheap_0 \) and \( \aheap_1 \) region-by-region
  using a set of {\em local rules}:
  \begin{compactitem}
  \item {\bf (Decomposition)} Suppose $\aheap_0$ and $\aheap_1$ can be
    decomposed as \( \aheap_0 = \aheap_{0,0} \lsep \aheap_{0,1} \) and
    \( \aheap_1 = \smash{\aheap_{1,0} \lsep \aheap_{1,1}} \).  And if the
    corresponding sub-regions can be shown to satisfy the inclusions
    \[
    \compareheap( \relcw, \aheap_{0,0}, \aheap_{1,0} ) = (\true,
    \relcw') \quad\text{and}\quad \compareheap( \relcw', \aheap_{0,1},
    \aheap_{1,0} ) = (\true, \relcw'') \;,\]
    then the overall inclusion
    holds---\( \compareheap( \relcw, \aheap_0, \aheap_1 ) \) returns
    \( (\true, \relcw'') \);
  \item {\bf (Points-to edges)}
    If \( \aheap_0 = \alpha_0 \cdot \fldf \pt \beta_0 \lsep \aheap_{0,r} \),
    \( \aheap_1 = \alpha_1 \cdot \fldf \pt \beta_1 \lsep \aheap_{1,r} \) and
    \( \relcw( \alpha_1 ) = \alpha_0 \), then we can conclude inclusion
    holds locally and extend \( \relcw \) with \( \relcw( \beta_1 ) =
    \beta_0 \);
  \item {\bf (Unfolding)}
    If there is an unfolding of $\aheap_1$ called $\aheap_{1,u}$ such
    that \( \compareheap( \Phi, \aheap_0, \aheap_{1,u} ) = (\true, \relcw') \),
    then \( \compareheap( \relcw,
    \smash{\aheap_0, \aheap_1} ) =( \true, \relcw') \).
  \end{compactitem}
\item When \( \compareheap( \relcw, \aheap_0, \aheap_1 ) \) succeeds and
  returns \( (\true, \relcw') \), it means the inclusion holds with
  respect to the shape.  We, however,
  still need to check for inclusion with respect to the numeric
  properties.
  Recall that the base numeric domain elements $\anum_0 \in
  \adomnump{\gvars{\aheap_0}}$ and $\anum_1 \in
  \adomnump{\gvars{\aheap_1}}$ have incomparable sets of symbolic
  variables.  An inclusion check in the base numeric domain
  can only be performed after renaming symbolic
  names so that they are consistent.  The node mapping $\relcw'$
  computed by the above is precisely the renaming that is needed.
  Thus, the last step to perform to decide inclusion is to compute
  \( \comparenum( \anum_0, \renamenum( \relcw', \anum_1 ) ) \) and
  return it as a result for \( \comparecomb( \relcw,
  \smash{(\aheap_0, \anum_0), (\aheap_1, \anum_1))} \).
  Note that function \( \renamenum \) should be sound in the following
  sense:
  \[
  \forall \anum \in \adomnump{V}, \;
  \forall \valua \in \gammanump{V}( \anum ), \;
  (\valua \circ \relcw) \in
  \gammanump{\relcw(V)}( \renamenum( \relcw, \anum ) )
  \]
  where $\relcw(V)$ is the set of symbolic variables obtained by
  applying $\relcw$ to set $V$.
\item If any of the above steps fail, \( \comparemem \) returns \( \false \).
\end{compactenum}

To summarize, the soundness conditions of the inclusion tests for the
lower-level domains on which $\comparemem$ relies are as follows:
\begin{condition}[Soundness of inclusion tests]
\[
\begin{array}{ll}
  1. &
  \text{If}\;
  \comparenum( \anum_0, \anum_1 ) = \true
  \;,\text{then}\;
  \gammanump{V}( \anum_0 ) \subseteq \gammanump{V}( \anum_1 )
  \;.
  \\[0.5ex]
  2. &
  \text{If}\;
      \compareheap( \relcw, \aheap_0, \aheap_1 ) = (\true, \relcw')
      %
      \;,\text{then}\;
      (\heap, \valua \circ \relcw') \in \gammaheap( \aheap_1 )
      \;\text{for all}\;
      (\heap, \valua) \in \gammaheap( \aheap_0 )
      \;.
  \\[0.5ex]
  3. &
  \text{If}\;
      \comparecomb( \relcw, (\aheap_0, \anum_0), (\aheap_1, \anum_1) )
      = (\true, \relcw')
      \;,\text{then}\;
      (\heap, \valua \circ \relcw') \in \gammacomb( \aheap_1, \anum_1 )
      \\&
      \text{for all}\;
      (\heap, \valua) \in \gammaheap( \aheap_0, \anum_0 )
      \;.
  \\
\end{array}
\]
\end{condition}

Returning to the
example in Figure~\ref{fig::cw},
after starting with \( \relcw = [\alpha'_0 \mapsto \alpha_0, \alpha'_1
\mapsto \alpha_1] \), the \( \compareheap \) operation consumes the points-to edges
one-by-one extending \( \relcw \) incrementally, unfolding the
inductive edges in the right argument before concluding that inclusion
holds in the shape domain.  With the final mapping
$\relcw'(\alpha'_i) = \alpha_i$ for all $i$,
the numeric inclusion simply needs to check that
\(
\comparenum( \alpha_3 \leq \alpha_5 \wedge \alpha_5 \leq \alpha_7,
\renamenum(\relcw', \alpha'_3 \leq \alpha'_5) )
=
\comparenum( \alpha_3 \leq \alpha_5 \wedge \alpha_5 \leq \alpha_7,
\alpha_3 \leq \alpha_5)
=
\true
\).

\subsection{Join and widening}
\label{sec:4:5:join}

As is standard, the \( \joinmem \) operation should satisfy the
following:
\begin{condition}[Soundness of $\joinmem$]
For all $\amem_0$ and $\amem_1$,
$\gammamem( \amem_0 ) \cup \gammamem( \amem_1)
\subseteq
\gammamem( \joinmem( \amem_0, \amem_1 ) )$.
\end{condition}

Like the comparison operator, the join operator takes two 
abstract heaps that have distinct sets of symbolic variables as input.
Additionally, it generates a new abstract heap, which requires another
set of symbolic variables, as it may not be possible to use the same
set as either input.
%
\begin{figure}[t]\centering
    \includegraphics[scale=1]{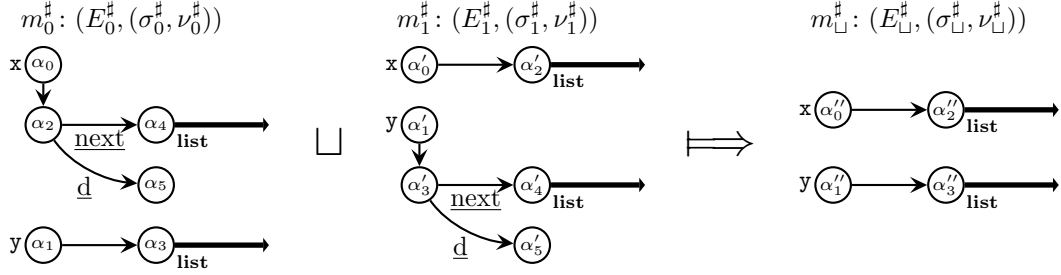}
  \caption{An abstract join showing the need for different sets of
    symbolic variables for each of the inputs and the result. The
    inputs are the two possible abstract heaps where
    a possibly-empty and a non-empty list are pointed to by
    two non-aliased program variables $\varx$ and $\vary$, so
    the most precise over-approximation is the abstract heap where
    both $\varx$ and $\vary$ point to two possibly-empty lists.}
  \label{fig::join}
\end{figure}
The example shown in Figure~\ref{fig::join} illustrates this situation.
In left input \( \smash{\amem_0} \), variable \( \varx \) points to a
non-empty list and \( \vary \) points to a possibly empty list, whereas
right input \( \smash{\amem_1} \) describes the opposite.
The most precise over-approximation of \( \smash{\amem_0} \) and
\( \smash{\amem_1} \) corresponds to the case where both \( \varx \) and
\( \vary \) point to lists of any length (as shown on the right side
of the figure).
These three elements all have distinct sets of nodes (that cannot be put
in a bijection).
Thus, the join algorithm uses a slightly different notion of symbolic
node mapping \( \reljoin \) that binds three-tuples of nodes
consisting of one node from each parameter and one node in the output
abstract heap.  Conceptually, the output abstract heap is a kind of
product construction, so it is composed of
new symbolic variables corresponding to pairs of nodes with 
one from each input.

Overall, the join algorithm proceeds in a similar way as the inclusion
test: the abstract heap join produces a mapping relating symbolic
variables along with a new abstract heap.  This mapping is then used
to rename symbolic variables in the base numeric domain elements
consistently to then apply the join in the base domain.  Similar to
the inclusion test, an initial mapping $\reljoin$ is constructed using
the abstract environment at the $\amems$ level and then extended
step-by-step at the $\aheaps$ level.
For instance, in Figure~\ref{fig::join}, the initial mapping is \( \{ (\alpha_0,
\alpha'_0, \alpha''_0), (\alpha_1, \alpha'_1, \alpha''_1) \} \), and then
pairs \( (\alpha_2, \alpha'_2, \alpha''_2) \) and  \( (\alpha_3, \alpha'_3,
\alpha''_3) \) are added by \( \joinheap \).
Note that nodes \( \alpha_3, \alpha_4, \alpha'_3, \alpha'_4 \) have no
counterpart in the result.

The local rules abstract heap join rules used in \( \joinheap \) belong
to two main categories:
\begin{compactitem}
\item {\bf (Bijection)}
  When two fragments of each input are isomorphic modulo \( \reljoin \),
  they can be joined into another such fragment.
  In the example, the points-to edges \( \alpha_0 \pt \alpha_2 \) and \( \alpha'_0
  \pt \alpha'_2 \) can both be over-approximated by \( \alpha''_0 \pt
  \alpha''_2 \).  Applying this rule adds the triple
  \( (\alpha_2, \alpha'_2, \alpha''_2) \) to the mapping \( \reljoin \).
\item {\bf (Weakening)}
  When a heap fragment can be shown to be included in
  a more simple, summary fragment (in terms of their concretizations),
  we can over-approximate the original fragment with the summary.
  For instance, fragment \( \alpha_2 \cdot \fldn \pt \alpha_3 \lsep
  \alpha_2 \cdot \fldd \pt \alpha_4 \lsep \icallpz{\alpha_3}{\indlist} \)
  can be shown to be included in
  \( \icallpz{\alpha_2}{\indlist} \).  The other input can be an
  effective means for directing the choice of possible
  summary fragments~\cite{xisa:sas:07,xisa:popl:08}.
\end{compactitem}
The widening operator \( \widenmem \) can be defined similarly to \( \joinmem \).
If the heap join rules enforce termination (\ie, \( \joinheap \) can
be used as a widening) and \( \joinnum \) is replaced
with a widening operator \( \widennum \), the cofibered domain definition
guarantees the resulting operator enforces termination~\cite{venet:sas:96}.

\subsection{Disjunctive abstract domain interface}

\begin{wrapfigure}{r}{0pt}\small
  \(
  \begin{array}{rlcl}
    \partitiondisj:
    & \ctxts \times \partsoffin{\adomdisj}
    & \longrightarrow
    & \adomdisj
    \\
    \collapsedisj:
    & \ctxts \times \adomdisj
    & \longrightarrow
    & \adomdisj
    \\
    \assigndisj:
    & \ctxts \times \lvals{\svars} \times \exprs{\svars} \times \adomdisj
    & \longrightarrow
    & \adomdisj
    \\
    \guarddisj:
    & \ctxts \times \exprs{\svars} \times \adomdisj
    & \longrightarrow
    & \adomdisj
    \\
    \allocdisj:
    & \ctxts \times \lvals{\svars} \times \bbN \times \adomdisj
    & \longrightarrow
    & \adomdisj
    \\
    \freedisj:
    & \ctxts \times \lvals{\svars} \times \bbN \times \adomdisj
    & \longrightarrow
    & \adomdisj
    \\
    \comparedisj:
    & \adomdisj \times \adomdisj
    & \longrightarrow
    & \sbools
    \\
    \joindisj:
    & \adomdisj \times \adomdisj
    & \longrightarrow
    & \adomdisj
    \\
    \widendisj:
    & \adomdisj \times \adomdisj
    & \longrightarrow
    & \adomdisj
    \\
  \end{array}
  \)
  \caption{Disjunctive abstraction interface.}
  \label{fig::tp}
\end{wrapfigure}
Recall from Sections~\ref{sec:3:3:summarize}
and~\ref{sec:4:2:unfold} that unfolding returns a finite set of
abstract elements interpreted disjunctively and thus justifies the
need for a disjunctive abstraction layer---independent of other possible
reasons like a desire for path-sensitivity.  In this subsection, we
describe the interface for a disjunctive abstraction layer $\adomdisj$
shown in Figure~\ref{fig::tp} that sits
above the memory layer $\amems$.  The following discussion
completes the picture of the abstract domain interfaces (cf.,
Figure~\ref{fig::domstruct}).
%
There are two main differences in the interface as compared to the one
for \( \adommem \).
First, the disjunctive abstract domain should provide two additional
operations \( \partitiondisj \) and \( \collapsedisj \) that
create and collapse partitions, respectively.  A partition represents
a disjunctive set of base domain elements.
Second, the transfer functions take an additional context information
parameter
\( \ctxt \in \ctxts \) that can be used in \( \adomdisj \) to tag each
disjunct with how it arose in the course of the abstract
interpretation.
\begin{condition}[Soundness of $\partitiondisj$ and $\collapsedisj$]
Let $\astate : \adomdisj$ and $\aState : \partsoffin{\adomdisj}$.
\[
\begin{array}{c@{\qquad\qquad}c}
  \bigcup \{ \gammadisj( \astate ) \mid \astate \in \aState \}
  \subseteq
  \gammadisj( \partitiondisj( \ctxt, \aState ) )
  &
  \gammadisj( \astate )
  \subseteq
  \gammadisj( \collapsedisj( \ctxt, \astate ) )
\end{array}
\]
\end{condition}
\noindent
Note that contexts play no role in the concretization, but operations
can use them, for example, to decide which disjuncts to merge
using \( \joinmem \) and which disjuncts to preserve.

%
Transfer functions \( \assigndisj \), \(\guarddisj\), \(\allocdisj\),
and \(\freedisj \)
all follow the same structure.
They first call the underlying operation on the memory abstract domain \( \adommem \) and then
apply the \( \partitiondisj \) partition on the output.
For instance, \( \assigndisj \) is defined as follows while
satisfying the expected soundness condition:
\[
\begin{array}{l@{\qquad}l}
  \assigndisj( \ctxt, \lval, \expr ) \defeq
  \partitiondisj( \ctxt, \{ \assignmem( \lval, \expr, \amem ) \mid
  \amem \in \astate \} )
  & \text{(definition)}
  \\[0.5ex]
  (\env, \subst{\heap}{\lvalsem{\lval}( \env, \heap)}{\exprsem{\expr}( \env, \heap )})
  \in \gammadisj( \assigndisj( \ctxt, \lval, \expr, \astate ) )
  & \text{(soundness)}
\end{array}
\]
Inclusion \((\comparedisj)\), join, and widening
operations should satisfy the
usual soundness conditions.
The \(\collapsedisj\) operator may be used to avoid generating too many
disjuncts (and termination of the analysis).

\section{A compositional abstract interpreter}
\label{sec:5:ai}
In this section, we assemble an abstract interpreter for the language
defined in Section~\ref{sec:2:conc} using the abstraction set up in
Section~\ref{sec:3:abs} and the interface of abstract operations
described in
Section~\ref{sec:4:ops}.

The abstract semantics of a program \( \prog \) is a function
\( \asem{\prog}: \adomdisj \rightarrow \adomdisj \), which takes
an abstract pre-condition as input and produces an abstract
post-condition as output.
Based on an abstract interpretation of the denotational semantics
of programs~\cite{ds:den:86,ds:den:09}, we can define the abstract
semantics by induction
over the syntax of programs as shown in Figure~\ref{fig::ai} in a
completely standard manner.
\begin{figure}
  \[
  \begin{array}{@{}c@{}}
  \begin{array}{r@{\;}c@{\;}l@{\quad}r@{\;}c@{\;}l}
    \asem{\prog_0; \prog_1}( \astate )
    & \defeq &
    \asem{\prog_1} \circ \asem{\prog_0}( \astate )
    &
    \asem{\ctrl: \lval = \cmalloc( n )}( \astate )
    & \defeq &
    \allocdisj( \ctxtof{\ctrl}, \lval, n, \astate )
    \\
    \asem{\ctrl: \lval = \expr}( \astate )
    & \defeq &
    \assigndisj( \ctxtof{\ctrl}, \lval, \expr, \astate )
    &
    \asem{\ctrl: \cfree( \lval )}( \astate )
    & \defeq &
    \freedisj( \ctxtof{\ctrl}, \lval, n, \astate )
    \\
  \end{array}
  \\[2.5ex]
  \begin{array}{rcl}
    \asem{\ctrl: \cif\;( \expr ) \; \prog_{\rm t} \; \celse \; \prog_{\rm f}}(
    \astate )
    & \defeq &
      \begin{array}[t]{@{}l@{}l}
        \joindisj( &\asem{\prog_{\rm t}}( \guarddisj( \ctxtof{\ctrl,\true},
        \expr, \astate ) ),
        \\
        &
        \asem{\prog_{\rm f}}( \guarddisj( \ctxtof{\ctrl,\false},
        \expr = \false, \astate ) ) )
      \end{array}
    \\[3.5ex]
    \asem{\ctrl: \cwhile\;( \expr ) \; \prog}( \astate )
    & \defeq &
    \guarddisj( \ctxtof{\ctrl,\false}, \expr = \false, \opalfp[{\astate}] \aF)
    \\
    & &
    \text{where}
      \begin{array}[t]{rlcl}
        \aF:
        & \adomdisj
        & \longrightarrow
        & \adomdisj
        \\
        & \astate_0
        & \longmapsto
        & \asem{\prog}( \guarddisj( \ctxtof{\ctrl,\true}, \expr, \astate_0 ) )
        \\
      \end{array}
    \\[-3ex]
  \end{array}
  \end{array}
  \]
  \caption{A denotational-style abstract interpreter for the
    programming language defined in
    Section~\ref{sec:2:1:concrete-semantics}.}
  \label{fig::ai}
\end{figure}
We let \( \ctxtof{\ldots} \) stand for computing some context
information based on, for example, the control state $\ell$ and/or the
branch taken.  This context information may
be used, for instance, by the disjunctive domain $\adomdisj$ to guide
trace partitioning~\cite{rm:toplas:07}.  The abstract transitions for
sequencing, assignment, dynamic memory allocation, and deallocation
are straightforward with the latter three calling the corresponding
transfer function in the top-layer abstract domain $\adomdisj$.  For
$\cif$, the pre-condition is first constrained by the guard condition
via $\guarddisj$ to interpret the two branches and then the resulting
states are joined via $\joindisj$.  For $\cwhile$, we write \( \alfp
\) for an abstract post--fixed-point operator.  The $\alfp$ operator
relies on \( \widendisj \) to terminate and on \( \comparedisj \) to
verify the stability of the abstract post-fixed point.  It may also
use \( \joindisj \) to increase the level of precision when computing
the first iterations. We omit  a full definition of \( \alfp \) as
there are many well-known
ways to obtain such an operator.  The most simple one consists of
applying only \( \widendisj \) until stabilization can be shown by
\( \comparedisj \).  We simply state its soundness condition:
\begin{condition}[Soundness of $\alfp$]
For all concrete transformers $F: \partsof{\sstates}
\rightarrow \partsof{\sstates}
\;\text{monotone}$, all abstract transformers
$\aF: \adomdisj \rightarrow \adomdisj$, and all abstract states
$\astate \in \adomdisj$,
\[
\text{if}\;
  F \circ \gammadisj \subseteq \gammadisj \circ \aF
\;,\text{then}\;
  \oplfp[{\gammadisj(\astate)}] F
  \subseteq
  \gammadisj( \opalfp[{\astate}] \aF )
  \;.
\]
\end{condition}
\noindent
We write $\oplfp[\States]$ for the least post--fixed point that is at
least $\States$ and similarly for $\opalfp[\astate]$.
Finally, the static analysis is sound in the following sense:
\begin{theorem}[Soundness of the analysis]
  \label{thm:sound}
  Let \( \prog \) be a program, and let \( \astate \in \adomdisj \) be an
  abstract pre-condition.
  Then, the result of the analysis is sound:
  \[
  \forall \state \in \gammadisj( \astate ), \;
  \semd{\prog}( \state ) \subseteq \gammadisj( \asem{\prog}( \astate ) )
  \;.
  \]
\end{theorem}
\noindent
Soundness can be proven by induction over the syntax of programs and by
composing the local soundness conditions of all abstract operators.

\relatedworkpara
An advantage of this iteration strategy, is that it leads
to an intuitive order of application of the abstract equations
corresponding to the program~\cite{cc:popl:77}, eliminating complex
iteration strategies~\cite{hor:iter:87}.
It also simplifies the choice of widening points~\cite{b:jfp:92}, as
it applies widening naturally, at loop heads, though it also allows
one to make different choices in strategy by, for example, modifying \( \alfp \) 
to unroll loop iterations~\cite{astree:pldi:03}.

\section{Conclusion}
\label{sec:6:conc}
We have presented a modular construction of a static analysis that is
able to reason both about the shape of data structures and their numeric
contents simultaneously.
Our construction is parametric in the desired numeric
abstraction, as well as the shape abstraction, making it possible to
continuously substitute improvements for each component or with
variants targeted at different classes of programs or even different programming
languages.
The main advantage of a modular construction is that it allows one to
design, prove, and implement each component of the analysis
independently.  Modular construction is a cornerstone of
quality software engineering, and our experience has been that this
nice property becomes even more important when dealing with the
complexity of creating a static analysis that simultaneously reasons about shape and numeric
properties.



\paragraph{Acknowledgments}

We are inspired by Dave Schmidt's continual effort in formalizing the
foundations of static analysis, abstract interpretation, and the
semantics of programming languages. 
We would also wish to thank Dave for being a pillar in the static
analysis community and supporting the research community with
often thankless, behind-the-scenes work.

This work has been supported in part by the European Research Council
under the FP7 grant agreement 278673, Project MemCAD and the United
States National Science Foundation under grant CCF-1055066.

\bibliographystyle{eptcs}
\bibliography{paper}
\end{document}